\pdfoutput=1
\documentclass[preprint]{aastex62}
\usepackage{booktabs}
\usepackage{multirow}

\newcommand{\CC}{{\tt C1.0+/0/24}}
\newcommand{\MM}{{\tt M1.0+/0/24}}
\newcommand{\Co}{{\tt C1.0}--{\tt C9.9}}
\newcommand{\Mo}{{\tt M1.0}--{\tt M9.9}}

\graphicspath{{./}{figures/}}

\received{November 11, 2019}
\revised{December 18, 2019}
\accepted{December 26, 2019}

\submitjournal{ApJ}

\shorttitle{Evaluating Consecutive-Day Forecasting Patterns}
\shortauthors{Park et al.}

\begin{document}

\title{A Comparison of Flare Forecasting Methods. IV. Evaluating Consecutive-Day Forecasting Patterns}

\correspondingauthor{Sung-Hong Park}
\email{shpark@isee.nagoya-u.ac.jp}

\author[0000-0001-9149-6547]{Sung-Hong Park}
\affiliation{Institute for Space-Earth Environmental Research, Nagoya University, Nagoya, Japan}

\author[0000-0003-0026-931X]{K. D. Leka}
\affiliation{Institute for Space-Earth Environmental Research, Nagoya University, Nagoya, Japan}
\affiliation{NorthWest Research Associates, Boulder, CO, USA}

\author[0000-0002-6814-6810]{Kanya Kusano}
\affiliation{Institute for Space-Earth Environmental Research, Nagoya University, Nagoya, Japan}

\author{Jesse Andries}
\affiliation{Solar-Terrestrial Center for Excellence, Royal Observatory of Belgium, Brussels, Belgium}

\author[0000-0003-3571-8728]{Graham Barnes}
\affiliation{NorthWest Research Associates, Boulder, CO, USA}

\author[0000-0002-6977-0885]{Suzy Bingham}
\affiliation{Met Office, Exeter, United Kingdom}

\author[0000-0002-4183-9895]{D. Shaun Bloomfield}
\affiliation{Northumbria University, Newcastle upon Tyne, United Kingdom}

\author[0000-0002-4830-9352]{Aoife E. McCloskey}
\affiliation{School of Physics, Trinity College Dublin, Dublin, Ireland}

\author[0000-0001-5307-8045]{Veronique Delouille}
\affiliation{Solar-Terrestrial Center for Excellence, Royal Observatory of Belgium, Brussels, Belgium}

\author{David Falconer}
\affiliation{NASA/NSSTC, Huntsville, AL, USA}

\author[0000-0001-9745-0400]{Peter T. Gallagher}
\affiliation{School of Cosmic Physics, Dublin Institute for Advanced Studies, Dublin, Ireland}

\author[0000-0001-6913-1330]{Manolis K. Georgoulis}
\affiliation{Department of Physics \& Astronomy, Georgia State University,Atlanta, GA, USA}
\affiliation{Research Center Astronomy and Applied Mathematics, Academy of Athens, Athens, Greece}

\author{Yuki Kubo}
\affiliation{National Institute of Information and Communications Technology, Tokyo, Japan}

\author[0000-0001-8969-9169]{Kangjin Lee}
\affiliation{Electronics and Telecommunications Research Institute, Daejeon, Republic of Korea}
\affiliation{School of Space Research, Kyung Hee University, Yongin, Republic of Korea}

\author{Sangwoo Lee}
\affiliation{SELab, Inc., Seoul, Republic of Korea}

\author[0000-0001-5655-9928]{Vasily Lobzin}
\affiliation{Space Weather Services, Bureau of Meteorology, Sydney, Australia}

\author{JunChul Mun}
\affiliation{Korean Space Weather Center, National Radio Research Agency, Jeju, Republic of Korea}

\author[0000-0002-9378-5315]{Sophie A. Murray}
\affiliation{School of Physics, Trinity College Dublin, Dublin, Ireland}
\affiliation{School of Cosmic Physics, Dublin Institute for Advanced Studies, Dublin, Ireland}

\author{Tarek A. M. Hamad Nageem}
\affiliation{University of Bradford, Bradford, United Kingdom}

\author[0000-0002-8637-1130]{Rami Qahwaji}
\affiliation{University of Bradford, Bradford, United Kingdom}

\author{Michael Sharpe}
\affiliation{Met Office, Exeter, United Kingdom}

\author[0000-0001-8123-4244]{Rob A. Steenburgh}
\affiliation{NOAA/NWS/NCEP Space Weather Prediction Center, Boulder, CO, USA}

\author[0000-0002-9176-2697]{Graham Steward}
\affiliation{Space Weather Services, Bureau of Meteorology, Sydney, Australia}

\author[0000-0002-6290-158X]{Michael Terkildsen}
\affiliation{Space Weather Services, Bureau of Meteorology, Sydney, Australia}

\begin{abstract}
A crucial challenge to successful flare prediction is forecasting periods that transition between ``flare-quiet'' and  ``flare-active''. Building on earlier studies in this series (Barnes {\it et al.} 2016; Leka {\it et al.} 2019a,b) in which we describe methodology, details, and results of flare forecasting comparison efforts, we focus here on patterns of forecast outcomes (success and failure) over multi-day periods. A novel analysis is developed to evaluate forecasting success in the context of catching the first event of flare-active periods, and conversely, of correctly predicting declining flare activity. We demonstrate these evaluation methods graphically and quantitatively as they provide both quick comparative evaluations and options for detailed analysis. For the testing interval 2016--2017, we determine the relative frequency distribution of two-day dichotomous forecast outcomes for three different event histories ({\it i.e.}, event/event, no-event/event and event/no-event), and use it to highlight performance differences between forecasting methods. A trend is identified across all forecasting methods that a high/low forecast probability on day-1 remains high/low on day-2 even though flaring activity is transitioning. For M-class and larger flares, we find that explicitly including persistence or prior flare history in computing forecasts helps to improve overall forecast performance. It is also found that using magnetic/modern data leads to improvement in catching the first-event/first-no-event transitions. Finally, 15\% of major ({\it i.e.}, M-class or above) flare days over the testing interval were effectively missed due to a lack of observations from instruments away from the Earth-Sun line.
\end{abstract}

\keywords{methods: statistical ---  Sun: activity --- Sun: flares}

\section{Introduction} 
\label{sec:intro}
Forecasting solar flares provides a laboratory with which to examine the understanding of these energetic events, but forecasts also serve to protect infrastructure impacted by our Sun's variable output on a daily basis. Success of forecasting, from both an empirical and physical point of view, has thus far been measured using statistical evaluations of correct forecasts with each event considered independently and equally. Both operationally and physically, however, it is crucial to understand the transitions from ``flare-quiet'' to ``flare-active'' and back again, as the Sun and its magnetic fields evolve, generating, storing, and finally releasing free magnetic energy in the form of energetic events.

A focused workshop on ``Benchmarks for Operational Solar Flare Forecasting Systems'' was held in 2017 at the Institute for Space-Earth Environmental Research (ISEE), Nagoya University. The primary objective of the workshop was to compare in a quantitative manner the performance characteristics of today's operational flare forecasting methods, as a follow-on to the ``All Clear'' workshop and its initial investigation into the methodology of forecast comparisons \citep[][Paper I]{2016ApJ...829...89B}. For this workshop, forecasts from 19 different operational flare forecasting methods were submitted for an agreed-upon testing interval of 1 January 2016 to 31 December 2017, following agreed-upon forecast intervals and event definitions, described in \citet[][Paper II]{2019ApJS..243...36L}. Results focusing on the head-to-head comparisons are presented there, using multiple evaluation methodologies including graphics and quantitative metrics based on both probabilistic and dichotomous forecasts as are often used for forecast validation \citep[][and references therein]{1976MWRv..104.1209W,2012ApJ...747L..41B,2016ApJ...829...89B,2017JSWSC...7A..20K,2017SpWea..15..577M}. Recognizing the small sample size and short testing period, it was found in Paper II that (1) many methods consistently demonstrate skill, although (2) no single method is ``best'' across multiple metrics, and (3) no method performs ``well'' ({\it i.e.}, better than 0.5 across numerous normalized skill scores and validation metrics, where 0.5 is halfway between no skill and perfect). Most importantly, the required methodology for providing fair and meaningful comparisons across forecasts was demonstrated -- centering primarily on common testing intervals, event definitions and evaluation using a variety of metrics. The question of {\it why} certain methods performed better or worse than others was examined in \citet[][Paper III]{2019ApJ...881..101L} by means of grouping the methods in different categories of their implementation details. In this context of broad implementation differences, the behavior and performance of the methods were evaluated. The results were weak due to both the non-uniqueness of the categorizations and the small sample size, but it was found that including prior flare history and active region evolution likely led to improved performance, with a further indication that including a human ``forecaster in the loop'' (FITL) was also advantageous.

During the workshop, the participants expressed interest in examining a particular interval in detail, {\it e.g.}, a case-study -- in part due to the fact that NOAA Active Region (AR) 12673 was fresh in our memories, having produced at least one flare greater than or equal to GOES M1.0 level each day for seven consecutive days from 4 September 2017. In a cursory manner, we found that many of the methods failed to predict a high chance of major flares for the first day of AR 12673's multi-day flaring activity. Yet while some methods subsequently and significantly increased their forecast probabilities on the second day so that they successfully predicted the second day of activity, other methods' forecasts showed little change for that second day ({\it i.e.}, the event day was missed again) regardless of the large flares that occurred on the previous day. These different behaviors between forecasting methods motivated us to explore forecast performance over consecutive days with variable event histories. 

Case studies are often used by operational facilities during forecaster training to target a particular known failure. Questions often asked in case-study examinations include, ``was the first flare (of a series over a multi-day period) predicted correctly?'' and ``did forecast probabilities in fact decrease as flare activity subsided?''. Such case studies can be misleading, however, as a method's performance during a particular interval may not reflect its performance when numerous different intervals are considered. Here we extend this line of questioning to examine particular patterns of forecasting behavior using a multi-day analysis; specifically, we examine sets of two consecutive days where at least one of those days includes an event. We test the hypothesis that including some aspect of prior behavior or temporal evolution results in forecasts that are able to better adjust for varying flare activity. To evaluate this hypothesis, we present a newly developed analysis methodology to quantitatively evaluate specific temporally-oriented performance characteristics of solar flare forecasts.

\section{Methodology}
\label{sec:methodology}
We describe here the input data from the participating methods, and the methodology employed for evaluating outcome patterns of daily forecasts over consecutive forecast days, {\it i.e.} in the context of the challenges of predicting the first flaring and flare-quiet days described above. The results are presented later in Section~\ref{sec:results}.

\subsection{Participating Flare Forecasting Methods}
\label{subsec:methods}
The participants of the ISEE workshop brought 19 operational flare forecasting methods for analysis. Among them are several methods that have been implemented as operational flare forecasting systems at space weather Regional Warning Centres (RWCs) as well as at research institutes. Note that while no human forecaster intervenes in the forecast output of any research-institution-based methods, in general there are experienced forecasters at RWCs who take into account the implemented method outputs and may adjust them prior to issuing their official forecasts. Details of all participating flare forecasting methods can be found in Paper II and references therein. As in Paper III, for reference we reproduce an abbreviated version in Appendix~\ref{sec:appendix_method_table} as Table~\ref{tbl:methods} that lists the methods, relevant publications, and monikers/acronyms used here. As is clear from the earlier papers and Table~\ref{tbl:methods} here, not all 19 methods are completely independent but in some cases consist of different implementations of the same general approach ({\it c.f.}, the four versions of MAG4).

Full-disk daily forecasts were submitted and processed such that two event definitions were used consistently: 24-hour validity periods, effectively zero-hour latencies, and lower limits of {\tt C1.0} and {\tt M1.0} in GOES flare class; these are referred to here as \CC\ and \MM, respectively. For the methods that did not produce such full-disk exceedance forecasts, we converted their region-based forecasts to full-disk forecasts (described in Appendix B.1 of Paper II) and when appropriate, combined category-limited ({\it i.e.}, \Co, \Mo\ and {\tt X1.0+}) forecasts using conditional probabilities to provide exceedance forecasts (discussed in Appendix B.2 of Paper II). The testing interval was 1 January 2016 to 31 December 2017, inclusive. For each method and each event definition, a binary (yes/no) event list is compiled from recorded GOES flares in the NOAA Edited Solar Event Lists\footnote{\url{ftp://ftp.swpc.noaa.gov/pub/warehouse}}. Most of the methods issue their forecasts at 00:00 UT, while SIDC issues at 12:30 UT and NICT issues at 06:00 UT; for the latter two methods, custom event lists were created for the most consistent comparison possible with the other methods.

Over the course of the 731 days in 2016--2017, there are 188 event-days (25.7\%) and 26 event-days (3.6\%) for \CC\ and \MM, respectively. In case of missing forecasts for a method, these days are filled with 0\% probability values; as discussed in Paper II this is detrimental to the performance evaluation, but is fair for operational purposes. Probabilistic forecasts can be converted into dichotomous forecasts by setting a threshold $P_\mathrm{th}$ above which a forecast is classified as forecasting positively for an event. In this study, two different values of $P_\mathrm{th}$ are used. As with Papers II and III, $P_\mathrm{th}$\,=\,0.5 is applied by default, but in addition we examine the impact on performance when the threshold reflects the testing interval climatology instead: $P_\mathrm{th}$\,=\,0.257 and 0.036 for \CC\ and \MM, respectively. Note that this event frequency is very low (as discussed in Paper II), reflecting the fact that our testing period occurs on the declining phase of a weak activity solar cycle. All submitted forecast probabilities as well as relevant codes for data processing and analysis are freely available \citep{ffc3_data} so that readers may explore metrics and the effects of varying $P_\mathrm{th}$ as desired.

\subsection{At-A-Glance Performance}
\label{subsec:intro_ataglance}
\begin{figure}[t!]
\centering
\includegraphics[width=0.72\textwidth]{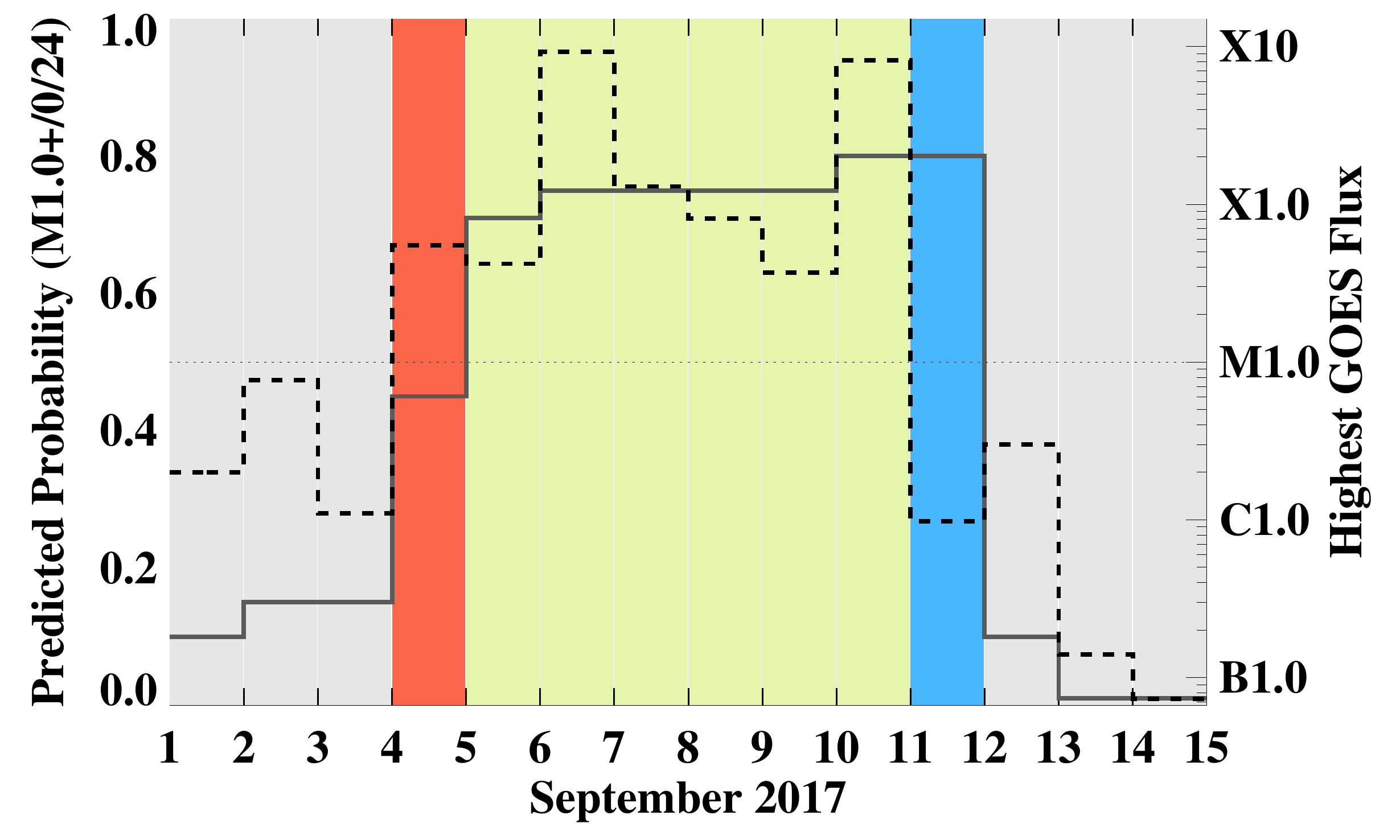}
\caption{An example of forecast probabilities (solid line) from an anonymous flare forecasting method for the \MM\ event definition is shown with the highest GOES soft X-ray flux (dashed line) observed each day over an interval of 1--15 September 2017. The GOES M1.0 level is marked with the horizontal dotted line for reference. Each day's forecast has a color-coded background shading that indicates one of the resulting dichotomous forecast outcomes using a $P_\mathrm{th}$\,=\,0.5 level: correct null (grey); miss (red); hit (green); false alarm (blue).}
\label{fig:prob_diagram_m_1_demo}
\end{figure}

We first investigate the overall performance of the daily flare forecasts from the 19 methods under consideration using a color-coded diagram of the forecasts in dichotomous form, as demonstrated in Figure~\ref{fig:prob_diagram_m_1_demo}. The diagram uses a designated $P_\mathrm{th}$ to color-code the dichotomous forecast outcomes ({\it i.e.}, hits, correct nulls, false alarms, misses), and the daily highest GOES Soft X-Ray flux is shown as well. The results are discussed in Section~\ref{subsec:results_at_a_glance}, below.

\subsection{Two-Day Analysis}
\label{subsec:intro_twoday}
Examining consecutive-day forecasting patterns enables us to begin a statistical analysis of that which is of interest in case studies. We consider three different ``event histories'' when one, or both, of the days includes an event (see Table~\ref{tbl:2day}); for simplicity we do not consider the no-event/no-event history. The provided forecasts for each of the two days then produces a ``forecast outcome pattern'' with four possible outcomes. That is, the two forecasts for the two-day period are considered as a unit (as opposed to each flare event being considered independently).

A goal here is to highlight mis-forecasting patterns in the context of the ``first event'' and the ``first quiet'' (effectively the ``last event'') of a flare-active period. In this context, the two event-history options are of an event occurring after a period of quiet (such as when a region begins to be flare active) during times of high but possibly varied flare activity, and of a no-event occurring after a flare-active period, meaning in this context the first flare-quiet day when activity is diminishing.

\begin{table}[t!]
    \centering
    \caption{Outcome Pattern Summary for Consecutive-Day Forecast Analysis}
    \label{tbl:2day}
    \begin{tabular}{c|c|c|c|c|c}
    \toprule
    Event History & if Forecast is:      & then Outcome is:       & \multirow{2}{*}{Label} & \multicolumn{2}{c}{\# Instances (\% out of the Total)} \\ 
    day-1 / day-2 & day-1 / day-2 & day-1 / day-2 &                        & \CC & \MM \\ \hline
    \multirow{4}{*}{Event / Event} & Yes / Yes & Hit / Hit & H-H & \multirow{4}{*}{121 (16.6\%)} & \multirow{4}{*}{12 (1.6\%)} \\
                    & Yes / No & Hit / Miss & H-M & & \\
                    & No / Yes & Miss / Hit & M-H & & \\
                    & No / No & Miss / Miss & M-M & & \\ \hline
    \multirow{4}{*}{No Event / Event} & Yes / Yes & False Alarm / Hit & F-H & \multirow{4}{*}{66 (9.0\%)} & \multirow{4}{*}{13 (1.8\%)} \\
                    & Yes / No & False Alarm / Miss & F-M  & & \\
                    & No / Yes & Correct Null / Hit & C-H & & \\
                    & No / No & Correct Null / Miss & C-M & & \\ \hline
    \multirow{4}{*}{Event / No Event} & Yes / Yes & Hit / False Alarm & H-F & \multirow{4}{*}{67 (9.2\%)} & \multirow{4}{*}{14 (1.9\%)} \\
                    & Yes / No & Hit / Correct Null & H-C & & \\
                    & No / Yes & Miss / False Alarm  & M-F & &  \\
                    & No / No & Miss / Correct Null & M-C & & \\
    \bottomrule
    \end{tabular}
\end{table}

\begin{figure}[h!]
\centering
\includegraphics[width=0.65\textwidth]{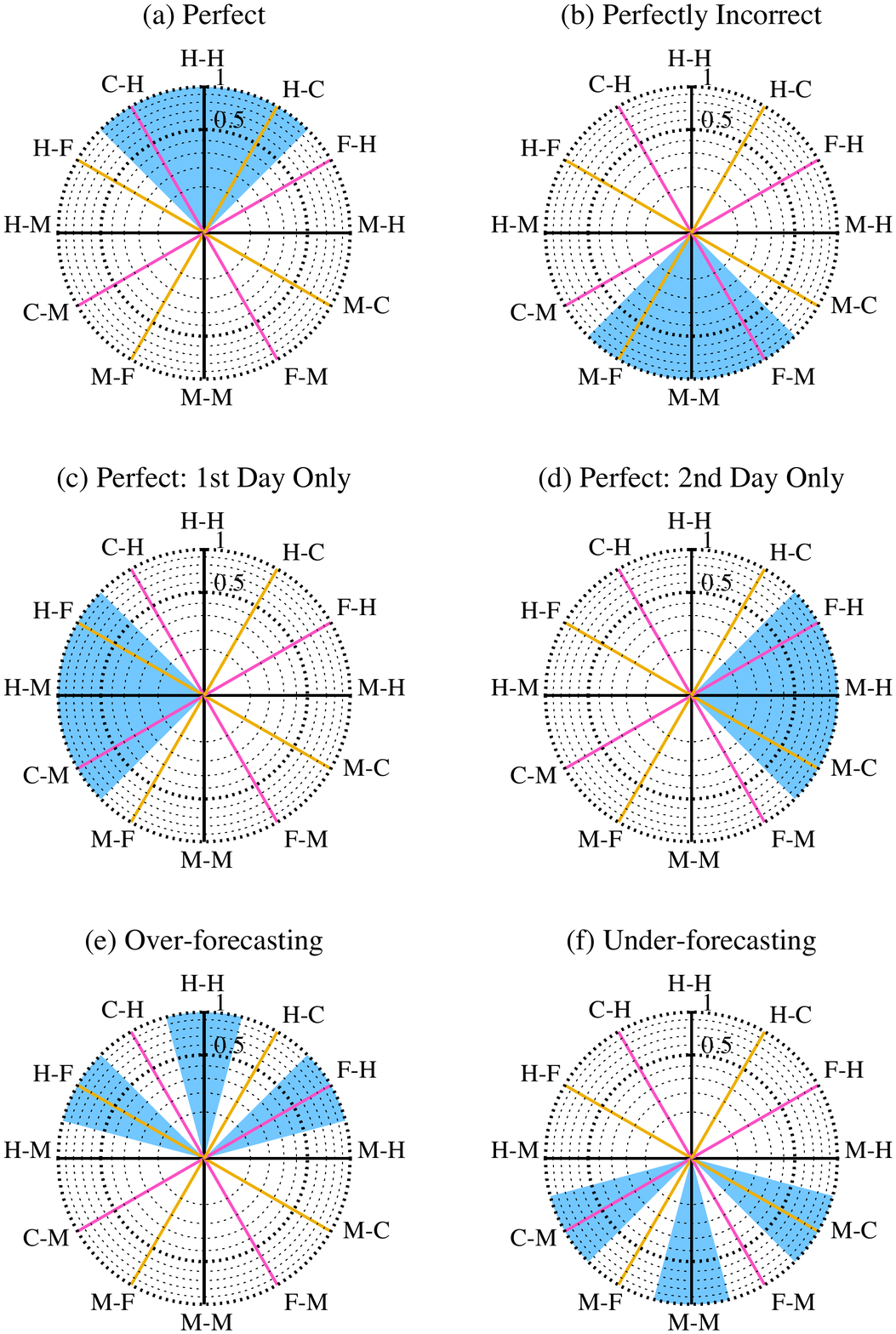}
\caption{Radar plots are used to demonstrate the relative frequency distribution of the two-day forecasting outcome patterns. Shown are idealized cases for (a) perfect, (b) perfectly incorrect, (c) perfect only on the first day, (d) perfect only on the second day, (e) consistently over-forecasting, and (f) consistently under-forecasting. The 2-letter axis labels indicate the direction for each of the 12 possible resulting outcome patterns ({\it i.e.}, four possible combinations for each of the three event histories, as described in Table~\ref{tbl:2day}). The extent of the colored wedge corresponds to the magnitude of the relative frequency indicated by the concentric dotted circles at intervals of 0.1. The radius of the concentric dotted circles is determined by the square root of the relative frequency, to emphasize the lower-frequency values which dominate in the present context. As an additional guide, the three event histories have color-coded axes: black (event/event), pink (no-event/event), and yellow (event/no-event). Note that the relative frequencies of the four combinations within each event history sum to unity.}
\label{fig:radar_demo}
\end{figure}

For this analysis we extend the graphical summary from the ``at-a-glance'' diagram in Figure~\ref{fig:prob_diagram_m_1_demo} to the categorizations in Table~\ref{tbl:2day}, and use a radar-plot format to summarize the performance characteristics of consecutive-day forecasts (Figure~\ref{fig:radar_demo}). For a specified $P_\mathrm{th}$ we compute the relative frequency with which a method's forecasts fall under each possible outcome pattern. For example, the number of occurrences of a particular outcome pattern ({\it e.g.}, the number of C-H outcomes) is divided by the total number of two-day forecasts within that particular event-day history (in this case no-event/event) over the 2-year interval (or 13 for \MM\ and 66 for \CC). Examples of radar-plots, and how they display particular outcomes (perfect, systematically over-forecasting, {\it etc.}) are shown in Figure~\ref{fig:radar_demo}. This presentation method statistically summarizes some of the important points of case studies.

For the specific question of how well the methods predict the first flare/first quiet (which is more explicitly the correct prediction of a change in activity from quiet to flare-active, and a change in activity from flare-active back to quiet) we can additionally specify which of the mixed-event outcome patterns have more or less impact on overall forecast performance. Successfully forecasting the first flaring day requires that at best, both the no-event and the following event day are correctly forecast; if only one of the two days is correctly forecast, it should at least be the event day rather than the no-event day. In other words, focusing on the no-event/event history, good forecasting performance dictates that the ``two-day-correct'' outcomes exceed the ``two-day-incorrect'' outcomes, and the ``first-day-incorrect/second-day-correct''outcomes exceed the ``first-day-correct/second-day-incorrect'' outcomes; using the labels, C-H\,$>$\,F-M, and F-H\,$>$\,C-M. For the radar plots this translates to asymmetric relative frequencies across particular $180^\circ$ sectors {\it i.e.} on an analog clock face: 11:00\,$>$\,05:00 and 02:00\,$>$\,08:00, respectively. Conversely, better performance forecasting the first flare-quiet day focuses on the event/no-event history, and requires H-C\,$>$\,M-F (01:00\,$>$\,07:00) and M-C\,$>$\,H-F (04:00\,$>$\,10:00).

\subsection{Two-Day Analysis plus Categorization}
\label{subsec:intro_twodayplus}
Finally we ask what implementation factors contribute to performance for consecutive-day forecasts. In this context we examine our original hypothesis that explicitly including temporal information in the forecasting method would improve performance -- including an improved ability to catch the first-event/first-no-event transitions. To this end, we focus on a few of the broad implementation options adopted in Paper III (see Table 5 in Paper III for the assignment of each forecasting method according to implementation option), and group the results by the outcome patterns. In some cases, three options as presented in Paper III are reduced to binary options here, as indicated, in order to maximize the sample size. The binary implementation options (hereafter referred to as ``BIOs'') that we focus on here: 

\paragraph*{Training Interval\,\,} describes a method's training data as ``\textit{Short}'', ``\textit{Long}'', or ``\textit{Hybrid}'' which are generally corresponding to {\it Solar Dynamics Observatory} \citep[SDO;][]{sdo} data only, multi-solar-cycle training, or a combination ({\it e.g.}, encompassing longer training, but then using {\it SDO} data for the forecasts themselves), respectively. For this work, we group the \textit{Short} and \textit{Hybrid} options together.

\paragraph*{Data Characterization\,\,} divides the methods into two broad groups, ``\textit{Simple}'' that rely on qualitative analysis or simpler inputs ({\it e.g.}, sunspot group categories) or ``\textit{Magnetic/Modern}'' quantitative analysis, most often comprising of photospheric magnetic field data.

\paragraph*{Persistence/Prior Flare Activity\,\,} describes whether a method qualitatively or quantitatively included persistence or prior flare activity in computing forecasts (or alternatively that no such information was included).

\paragraph*{Evolution\,\,} of underlying active regions is explicitly included in some methods, implicitly in others, but not at all for the majority of methods. Including evolution could take the form of, for example, tracking the evolution of sunspot group class and its impact on flaring rates \citep[as for MCEVOL, see][]{2016SoPh..291.1711M} or the contribution of a ``Forecaster in the Loop'' (FITL) judging a perceived change in a region's flaring rate and adjusting the forecast accordingly. \\

The BIOs of {\it Yes-Persistence} and {\it Yes-Evolution} explicitly include some aspect of time in the construction of the forecasts, while the other BIOs do not add any temporal dimension. This broad distinction is the focus of testing our above-stated hypothesis.

Following Paper III, we utilize a ``box and whisker'' presentation. However, instead of focusing on skill metrics, here we focus on the frequency of occurrence of the two-day forecast outcome patterns in the context of the three two-day event histories (event/event, no-event/event, and event/no-event).

\subsection{Targeted Questions}
\label{subsec:intro_questions}
The goal of this analysis is to investigate whether/how BIOs influence the forecast outcome patterns and performance results with respect to the event histories. Specifically, we examine the following questions:

\begin{enumerate}
	\item What is the impact of the BIOs on the 
	 independence of the two-day forecasts (meaning, does the forecast outcome for the first day 
	 significantly influence the forecast outcome for the second day)?	
	\item Is there any overall performance difference between BIOs within each particular categorization?
	\item Do any of the BIOs better predict both the first flare and first quiet?
	\item Do those BIOs that explicitly incorporate temporal information ({\it i.e.}, \textit{Yes-Persistence} and \textit{Yes-Evolution}) display performance differences as compared to those BIOs that do not include explicit temporal information?
\end{enumerate}

To address these questions, we analyze the results of the BIO performance by applying a variety of statistical methods to the forecast outcome patterns and their frequencies to answer the specific questions posed above. For example, when evaluating the influence (independence) of the first-day forecast outcome to the second-day forecast, we test the performance of the former in the context of the performance of the latter by evaluating the probability of rejecting the null hypothesis that the two are statistically independent. In contrast, when comparing the forecast performance across BIOs directly, we employ rank-sum tests since it is solely a comparative performance that is of interest. For the four questions here, the statistical approaches are described in detail in Appendix~\ref{sec:appendix_questions}. The data used for this analysis and for the box and whisker plots themselves are available \citep{ffc3_data}.

\section{Results}
\label{sec:results}
The analysis methods described above are applied to the forecasts for the participating methods, and the results are discussed below.

\subsection{At-A-Glance Performance Results}
\label{subsec:results_at_a_glance}
\begin{figure}[h!]
\centering
\includegraphics[width=0.76\textwidth]{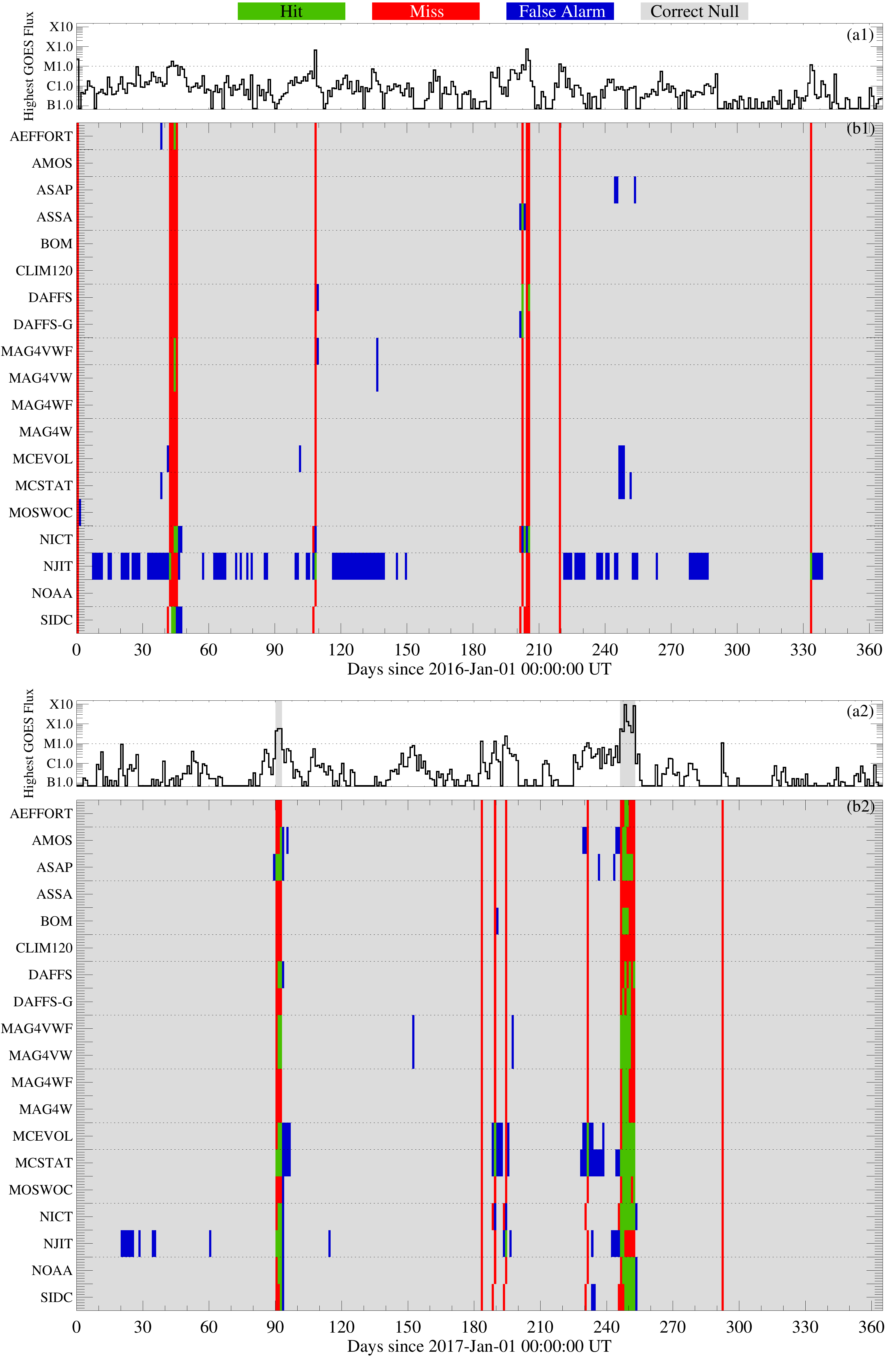}
\caption{The daily dichotomous forecast outcomes for \MM\ and $P_\mathrm{th}$\,=\,0.5 are shown over the two-year testing interval: 2016 (top panel) and 2017 (bottom panel). For both, panels (a1) and (a2) trace the highest GOES flux for each day, with the GOES {\tt M1.0} level indicated by a dotted line. Two intervals are marked with grey shading in panel (a2): 1--3 April 2017 and 4--10 September 2017, and discussed in Section~\ref{subsec:results_at_a_glance}. Panels (b1) and (b2) present the daily forecast outcomes -- hits (green), misses (red), false alarms (blue) and correct nulls (grey) -- by method, as labeled.}
\label{fig:overview_m}
\end{figure}

\begin{figure}[h!]
\centering
\includegraphics[width=0.76\textwidth]{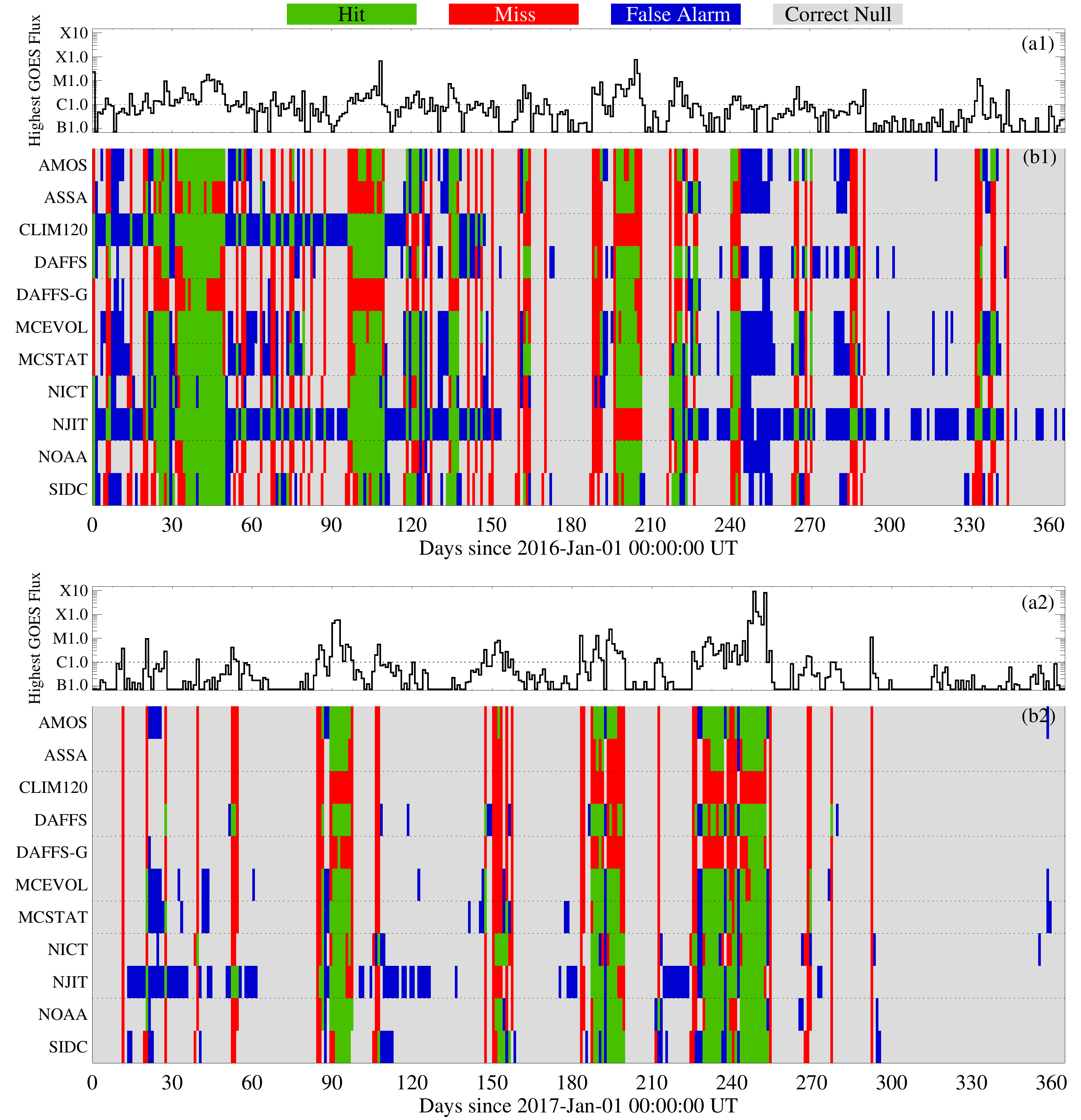}
\caption{Same as Figure~\ref{fig:overview_m}, but for daily \CC\ dichotomous forecasts with $P_\mathrm{th}$\,=\,0.5. The GOES {\tt C1.0} level is marked with the horizontal dotted lines in panels (a1) and (a2). Note that fewer methods produce forecasts for the \CC\ event definition.}
\label{fig:overview_c}
\end{figure}

Forecast outcome patterns begin to emerge when the time-series of forecasts is presented (Figures~\ref{fig:overview_m} and~\ref{fig:overview_c} {\it c.f.} Figure~\ref{fig:prob_diagram_m_1_demo})\footnote{Careful examination of Figures~\ref{fig:overview_m} and~\ref{fig:overview_c} reveals a slight offset in the temporal axis for SIDC and NICT where, {\it e.g.}, red-shaded missed event-days occasionally appear one day earlier than other methods. As mentioned in Section~\ref{subsec:methods}, custom event lists were created for these two methods due to their significantly different forecast issuance times. As discussed in Paper II, this will change the results slightly, but provides our best solution to the issue at hand.}. From the GOES traces of the highest daily soft X-ray flux (Figures~\ref{fig:overview_m} and~\ref{fig:overview_c}, top panels) it is obvious that, overall, activity is very low as these two years are toward the end of the solar activity cycle. Most methods successfully predict the long intervals of no activity ({\it i.e.}, grey-shaded ``correct nulls''), especially for the \MM\ definition. Most methods had some intervals of correct prediction (green) and all methods had missed events (red) for both event definitions.

However, even with $P_\mathrm{th}$\,=\,0.5, CLIM120 (the previous 120-day climatology forecast) shows many days of false alarms (blue) for \CC\ (as do MCSTAT and MCEVOL to a lesser degree), while NJIT shows many instances of false alarms for both event definitions. Note that CLIM120 with $P_\mathrm{th}$\,=\,0.5 was unable to make any correct \MM\ dichotomous forecasts over this testing interval, however this is expected at some level, since the climatological rate is well below the $P_\mathrm{th}$ value chosen. On the other hand, for the \CC\ definition, CLIM120 produced many false alarms because the climatological rate was higher than $P_\mathrm{th}$\,=\,0.5 for the first half of the testing interval, as discussed in Paper II.

Two intervals are highlighted in panel (a2) of Figure~\ref{fig:overview_m}: 1--3 April 2017 and 4--10 September 2017. These two time periods present patterns of interest for case studies, due to their distinct commencement, continuation, then cessation of flaring. For the start of flaring, a no-event-day is followed by an event-day and many methods correctly predict the former and miss the latter ({\it i.e.}, C-M). Moving farther into the 
flaring interval we have consecutive event-days; for any particular event/event history, some methods miss the first but succeed or hit for the second event-day (M-H) or {\it vice versa} (H-M) while others may miss both event-days (M-M), which is the worst forecast outcome. Finally approaching the end of the flaring interval, some methods correctly forecast the last event-day (the ``last flare'') and its following quiet day (H-C), but many follow a correct hit with a false alarm (H-F), thus not recognizing the cessation of flaring.

As discussed in Paper II some methods lack high-probability forecasts, especially for the larger-threshold \MM\ event definition. Many forecasts then register as misses for $P_\mathrm{th}$\,=\,0.5. We examine miss outcomes for larger events ({\it i.e.}, \MM) further in Section~\ref{subsec:results_twoday} in the context of $P_\mathrm{th}$\,=\,CLIM and still further in Section~\ref{subsec:limb_events}. Recognizing that the timelines of forecast outcomes presented here still essentially take the form of a case study (or two), we turn next to a statistical analysis of the trends.

\subsection{Two-day Analysis of Forecast Outcome Patterns}
\label{subsec:results_twoday}
\begin{figure}[h!]
\centering
\includegraphics[width=1\textwidth]{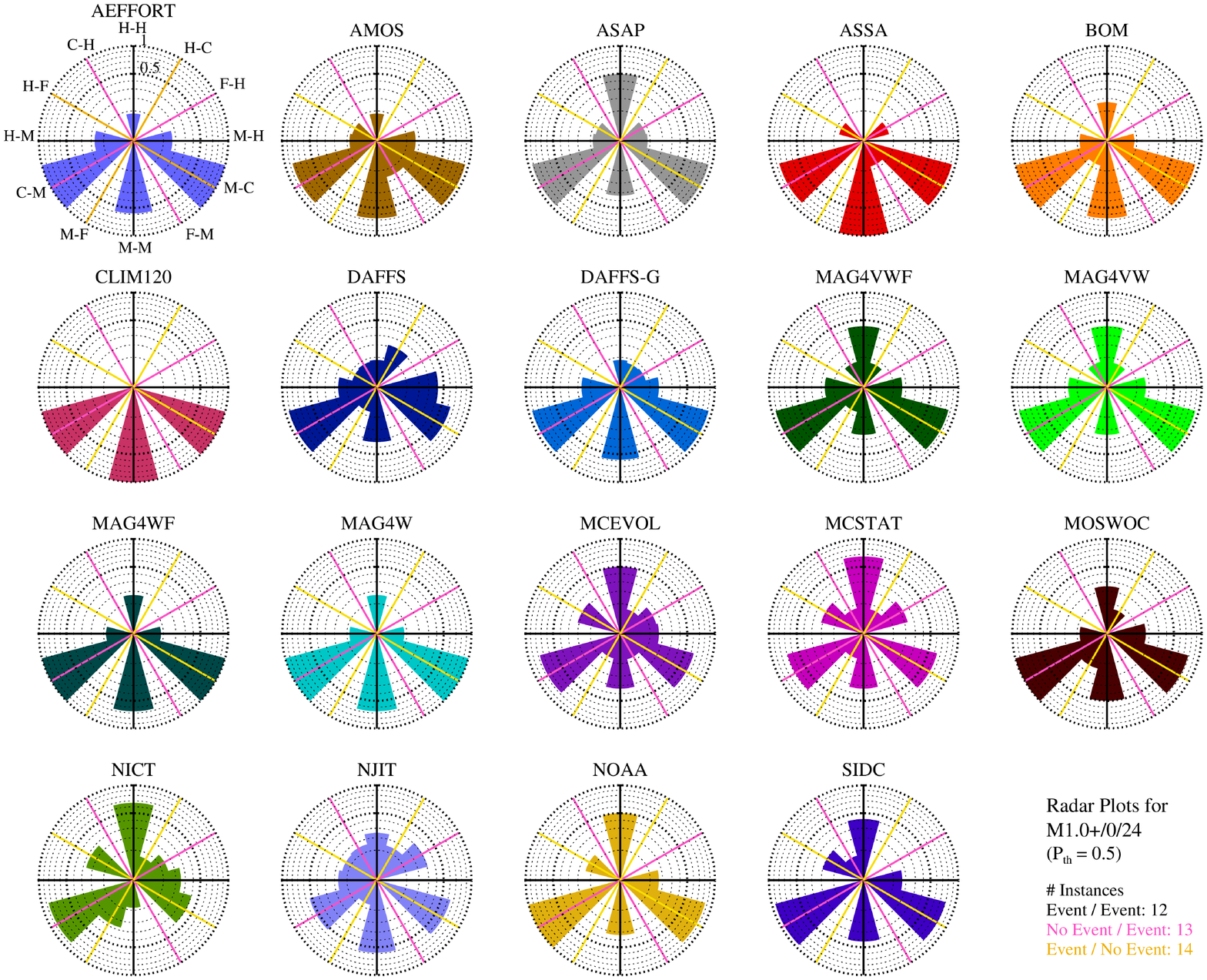}
\caption{Radar plots for each flare forecasting method, indicating the relative frequency distribution of the two-day forecasting patterns in the \MM\ dichotomous forecasts with $P_\mathrm{th}$\,=\,0.5 over the 2016--2017 testing interval. Colors for each method follow those used in Paper II.}
\label{fig:radar_m}
\end{figure}

\begin{figure}[h!]
\centering
\includegraphics[width=1\textwidth]{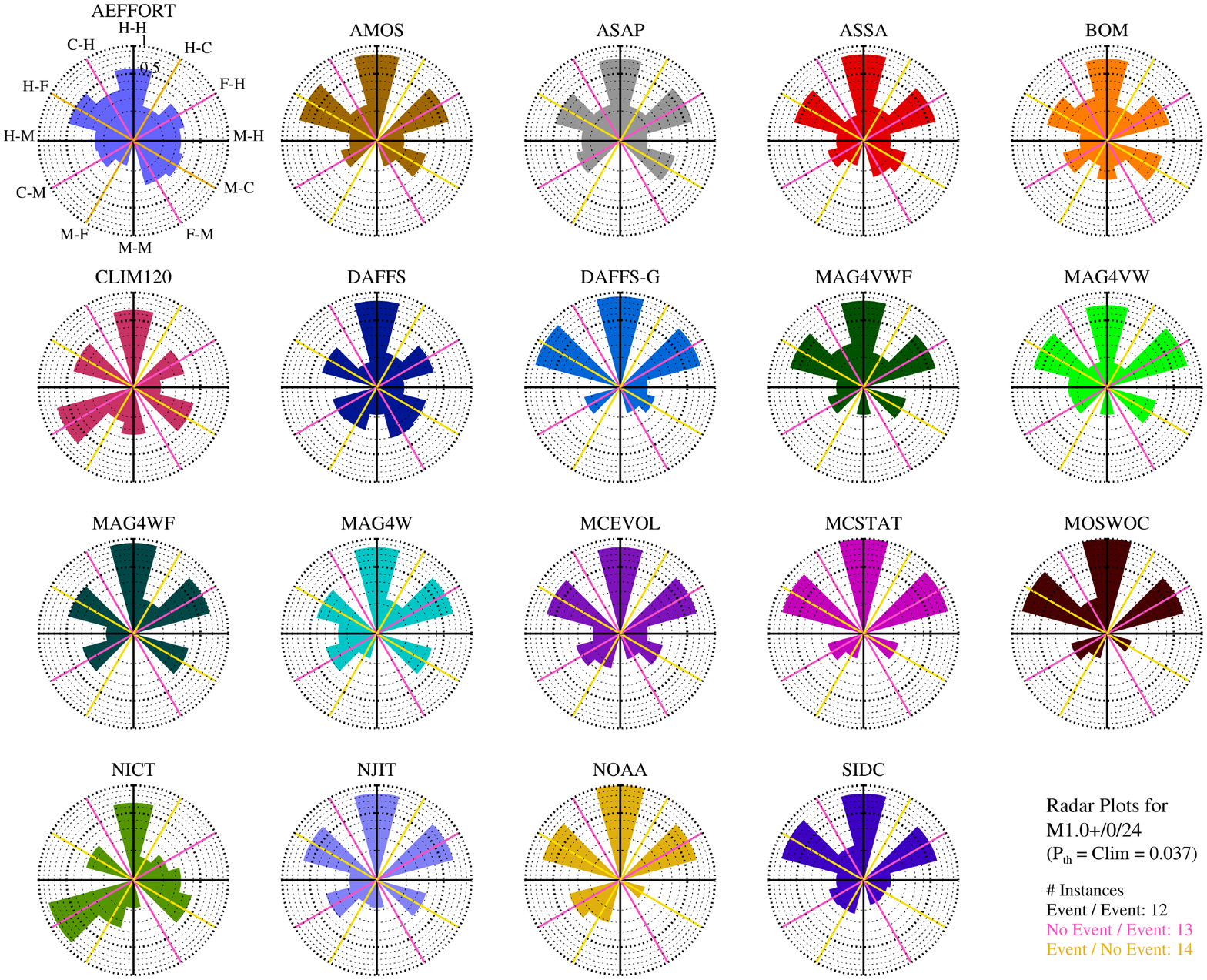}
\caption{Same as Figure~\ref{fig:radar_m}, but for \MM\ dichotomous forecasts with $P_\mathrm{th}$\,=\,CLIM, where CLIM refers to the climatological rate for the testing interval ({\it i.e.}, 0.036) of one or more flares occurring for the \MM\ event definition over the two-year testing interval.}
\label{fig:radar_m_clim}
\end{figure}

\begin{figure}[h!]
\centering
\includegraphics[width=1\textwidth]{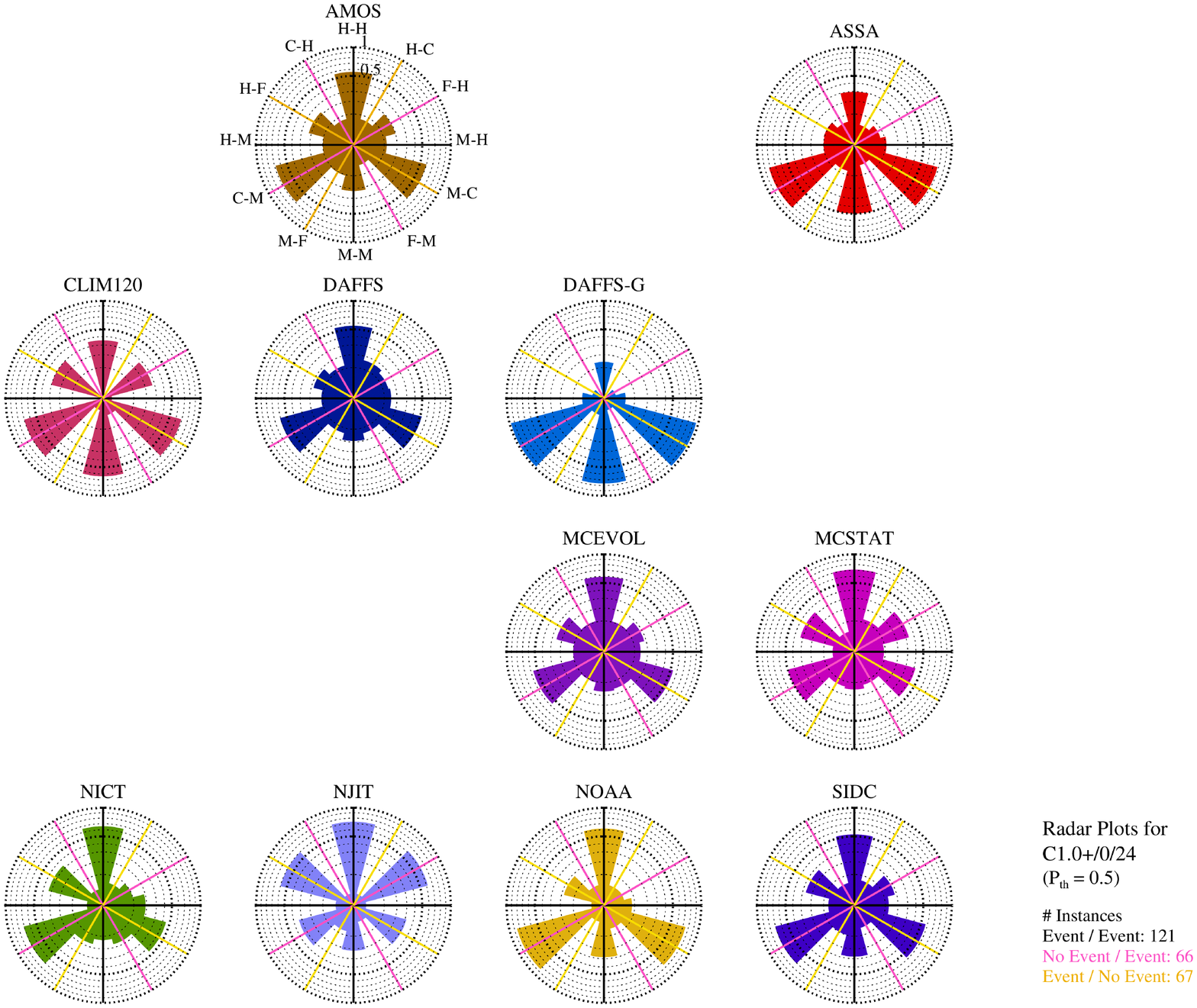}
\caption{Same as Figure~\ref{fig:radar_m}, but for the \CC\ dichotomous forecasts with $P_\mathrm{th}$\,=\,0.5.}
\label{fig:radar_c}
\end{figure}

\begin{figure}[h!]
\centering
\includegraphics[width=1\textwidth]{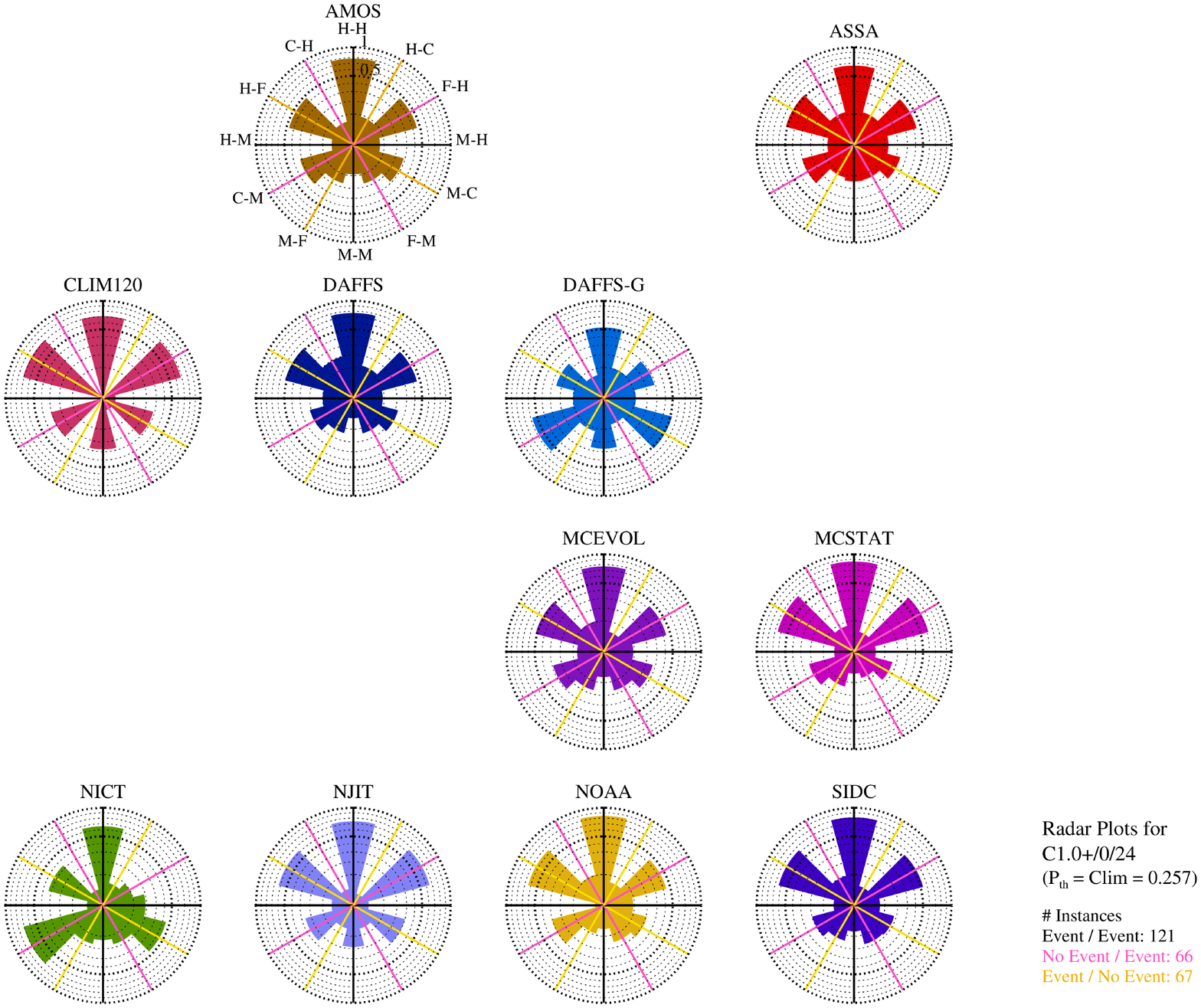}
\caption{Same as Figure~\ref{fig:radar_m}, but for the \CC\ dichotomous forecasts with $P_\mathrm{th}$\,=\,CLIM, where CLIM refers to the climatological event rate ({\it i.e.}, 0.257) for \CC\ over the testing interval.}
\label{fig:radar_c_clim}
\end{figure}

As described in Section~\ref{subsec:intro_twoday}, we use a radar plot format to analyze the performance of each forecasting method's outcome patterns, comparing the results between $P_\mathrm{th}$\,=\,0.5 and $P_\mathrm{th}$\,=\,CLIM as well as between the two event definitions \MM\ and \CC, respectively, in Figures~\ref{fig:radar_m}--\ref{fig:radar_c_clim}.

Referring back to Figure~\ref{fig:radar_demo}, for \MM\ and $P_\mathrm{th}$\,=\,0.5 (Figure~\ref{fig:radar_m}) we see a general trend of under-forecasting with a dominance of M-C and C-M outcomes ({\it i.e.}, corresponding to 04:00 and 08:00 sectors on an analog clock face, respectively) as the most frequent outcomes, with some methods additionally showing a high frequency of M-M. This analysis highlights the fact that for $P_\mathrm{th}$\,=\,0.5 most methods fail almost equally to forecast both the first flaring day of an increasing-activity period and the first flare-quiet ({\it i.e.}, second) day in a decreasing-activity period. That being said, while there is varying degree of event/event history failure (03:00, 06:00, 09:00 on the analog-clock analogy, respectively), some methods (ASAP, MAG4VWF, MAG4VW, MCEVOL, MCSTAT, NICT, NOAA) do show higher H-H frequency of success than M-M frequency of failure.

The performance changes dramatically with $P_\mathrm{th}$\,=\,CLIM\,=\,0.037 (Figure~\ref{fig:radar_m_clim}) which is a notably low $P_\mathrm{th}$. The trend is now over-forecasting, with most methods correctly forecasting both days for the event/event history, but with a high frequency of false alarms for the mixed-event histories, {\it i.e.}, a dominance of F-H and H-F (02:00 and 10:00 sectors, respectively). There are now almost no complete failures for the event/event history ({\it i.e.}, a low frequency of M-M) and a few methods with almost perfect H-H results (MCSTAT, MOSWOC, NOAA).

We next compare the radar plots for the \CC\ forecasts with $P_\mathrm{th}$\,=\,0.5 (Figure~\ref{fig:radar_c}) and with $P_\mathrm{th}$\,=\,CLIM = 0.257 (Figure~\ref{fig:radar_c_clim}); not all methods produce forecasts for this event definition, and following the approach in Paper II we leave the missing radar plots blank. The threshold probability difference between 0.5 and CLIM is not as extreme as for the \MM\ case, hence the results are impacted less by the change in $P_\mathrm{th}$. The same shift from under- to over-forecasting is seen, especially for the mixed-event histories.

Comparing \MM\ to \CC\ radar plots (for those methods which produce both), the majority of methods show a higher frequency of H-H than M-M for \CC\ which was rarely the case for \MM\ and $P_\mathrm{th}$\,=\,0.5. In both $P_\mathrm{th}$=0.5 and $P_\mathrm{th}$=CLIM and for both \CC\ and \MM\ there is a consistently very low frequency for the two-day-correct outcome patterns in the mixed-event histories (C-H and H-C in the 11:00 and 13:00 sectors, respectively).

The performance of predicting the first flaring day for \MM\ and $P_\mathrm{th}$\,=\,0.5 is, by all accounts, poor across all methods; only NJIT even weakly registers just one of the relevant two criteria described in Section~\ref{subsec:intro_twoday} ({\it i.e.}, C-H\,$>$\,F-M). For \MM\ and $P_\mathrm{th}$\,=\,CLIM, several methods register weak positive performance according to the criteria ({\it i.e.}, C-H\,$>$\,F-M and F-H\,$>$\,C-M). For \CC\ and $P_\mathrm{th}$\,=\,0.5, forecasting the first flaring day shows a modicum of success only for NJIT and only according to the second of the two criteria {\it i.e.}, F-H\,$>$\,C-M. However, for \CC\ and $P_\mathrm{th}$\,=\,CLIM the majority of methods can claim some success according to at least that second of the two criteria.

Forecasting the first quiet day (or the last flare) for \MM\ and $P_\mathrm{th}$\,=\,0.5 shows some promise for DAFFS, DAFFS-G, MAG4VM and NJIT by the analogous criteria ({\it i.e.}, H-C\,$>$\,M-F and M-C\,$>$\,H-F), while almost all methods have some success according to one of the criteria ({\it i.e.}, M-C\,$>$\,H-F), due to the prevalence of under-forecasting. For \MM\ and $P_\mathrm{th}$\,=\,CLIM, the situation changes: only NICT succeeds and only according to the second criteria M-C\,$>$\,H-F. Turning to the \CC\ definition, the results are similar as for \MM\ due to under-forecasting (for all methods except NJIT) at $P_\mathrm{th}$\,=\,0.5, while at $P_\mathrm{th}$\,=\,CLIM only DAFFS-G and NICT show some success according to M-C\,$>$\,H-F.

\subsection{Two-day Analysis plus Categorization: Results}
\label{subsec:results_twodayplus}
The goal of the next analysis is to investigate the two-day event histories and outcome patterns in the context of differences in BIOs that summarize specific method differences suspected of influencing the results. We turn with specific interest to results from the BIOs that include explicit temporal information ({\it i.e.}, \textit{Yes-Persistence} and \textit{Yes-Evolution}) as compared to those which include no temporal information.

\begin{figure}[h!]
\centering
\includegraphics[width=0.982\textwidth]{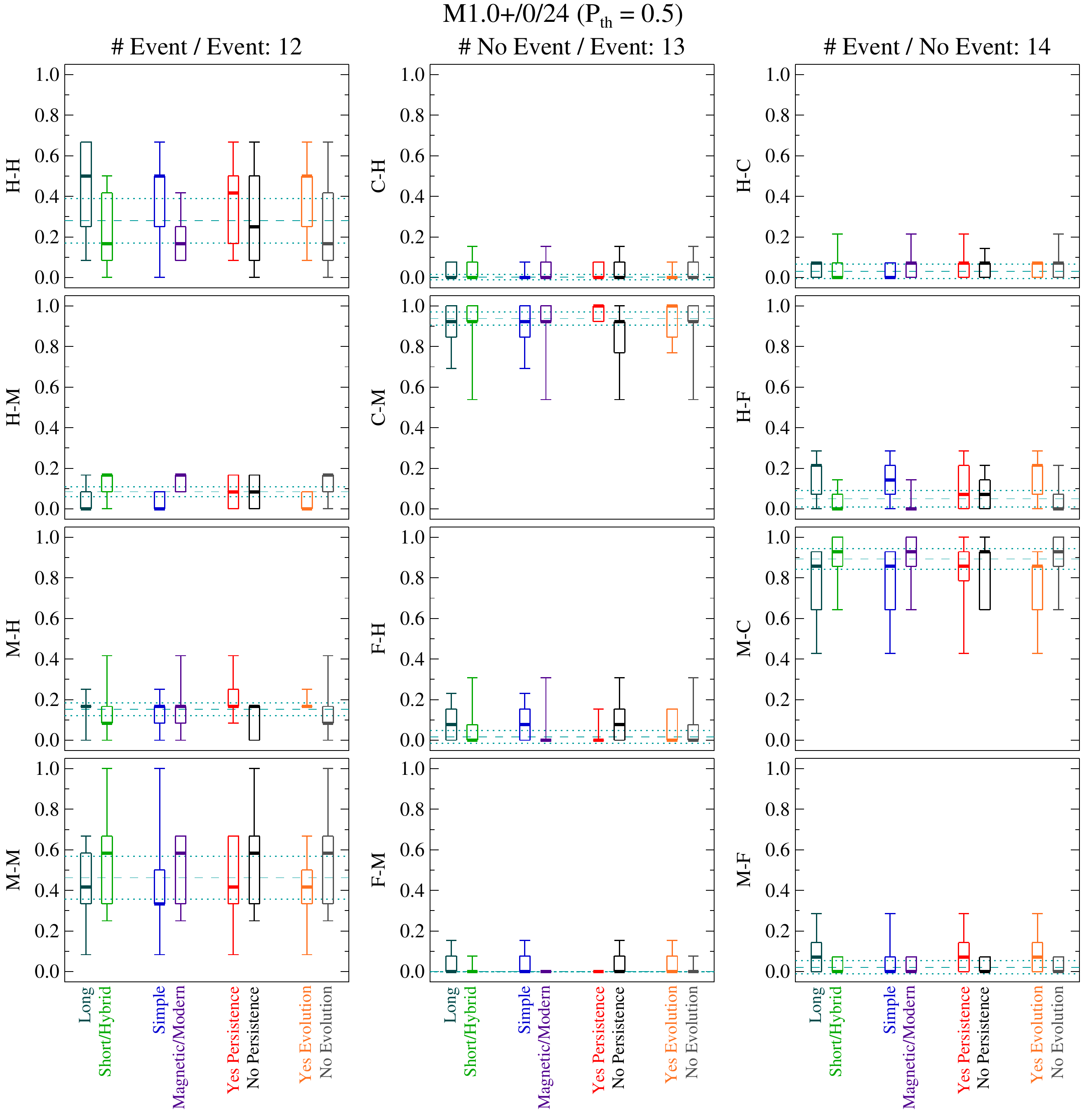}
\caption{Comparison of the relative frequencies of the two-day forecast outcome patterns in the \MM\ dichotomous forecasts with $P_\mathrm{th}$\,=\,0.5 between different groups of methods, as categorized according to the description in the text. Box and whisker plots display the 25th (lower edge) and 75th (upper edge) percentiles, the median (horizontal thick line inside the box), and the minimum and maximum of the sample (``whiskers''). Note that if the median coincides with either the 25th or 75th percentile, that box edge will be thicker. Also plotted are the mean (dashed) $\pm$ the standard deviation (dotted) of median values of relative frequencies from 100 sets of 9 randomly selected methods among all 18 methods except CLIM120. The top row for each of the three event histories is the ``correct/correct'' result, hence a higher frequency is better; all other rows indicate the frequency of results that include forecast errors, hence lower scores are better.}
\label{fig:box_whisker_m}
\end{figure}

\begin{figure}[h!]
\centering
\includegraphics[width=0.982\textwidth]{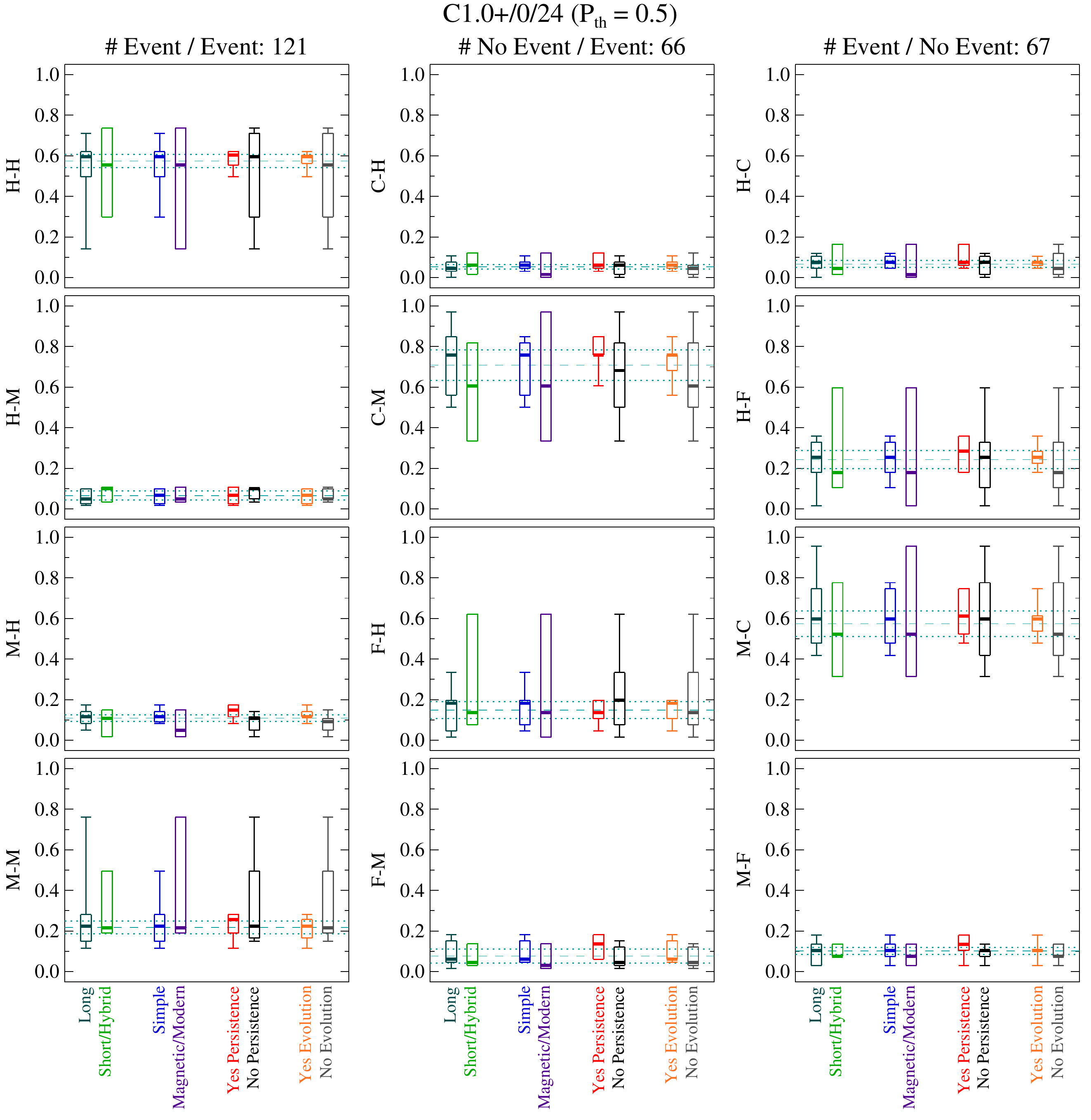}
\caption{Same as in Figure~\ref{fig:box_whisker_m}, but for the \CC\ dichotomous forecasts with $P_\mathrm{th}$\,=\,0.5. The mean (dashed) $\pm$ the standard deviation (dotted) of median values of relative frequencies are calculated from 100 sets of 5 randomly selected methods among all 10 available methods for \CC\ except CLIM120.}
\label{fig:box_whisker_c}
\end{figure}

\begin{figure}[h!]
\centering
\includegraphics[width=0.982\textwidth]{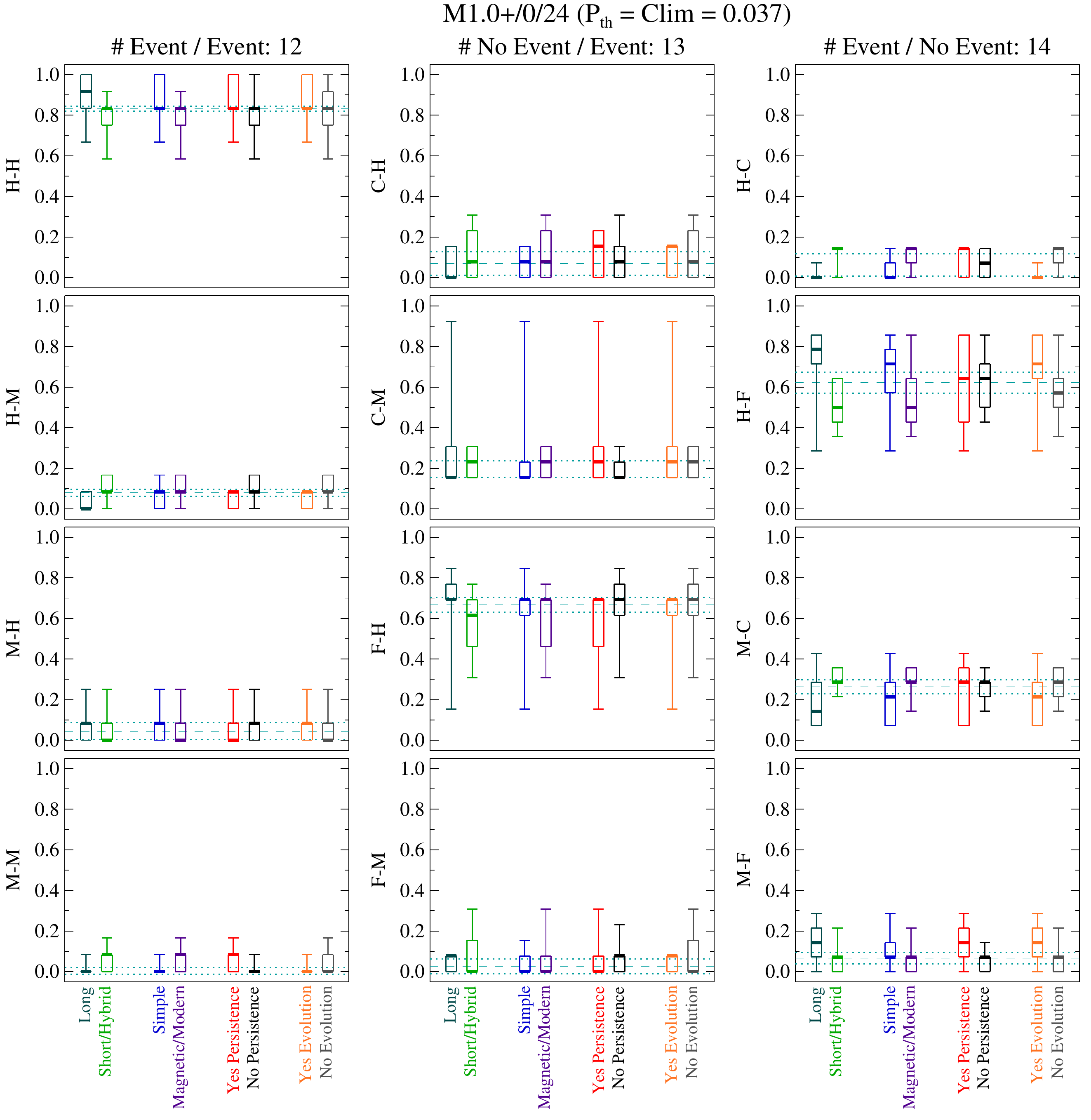}
\caption{Same as in Figure~\ref{fig:box_whisker_m}, but for the \MM\ dichotomous forecasts with $P_\mathrm{th}$\,=\,CLIM.}
\label{fig:box_whisker_m_clim}
\end{figure}

\begin{figure}[h!]
\centering
\includegraphics[width=0.983\textwidth]{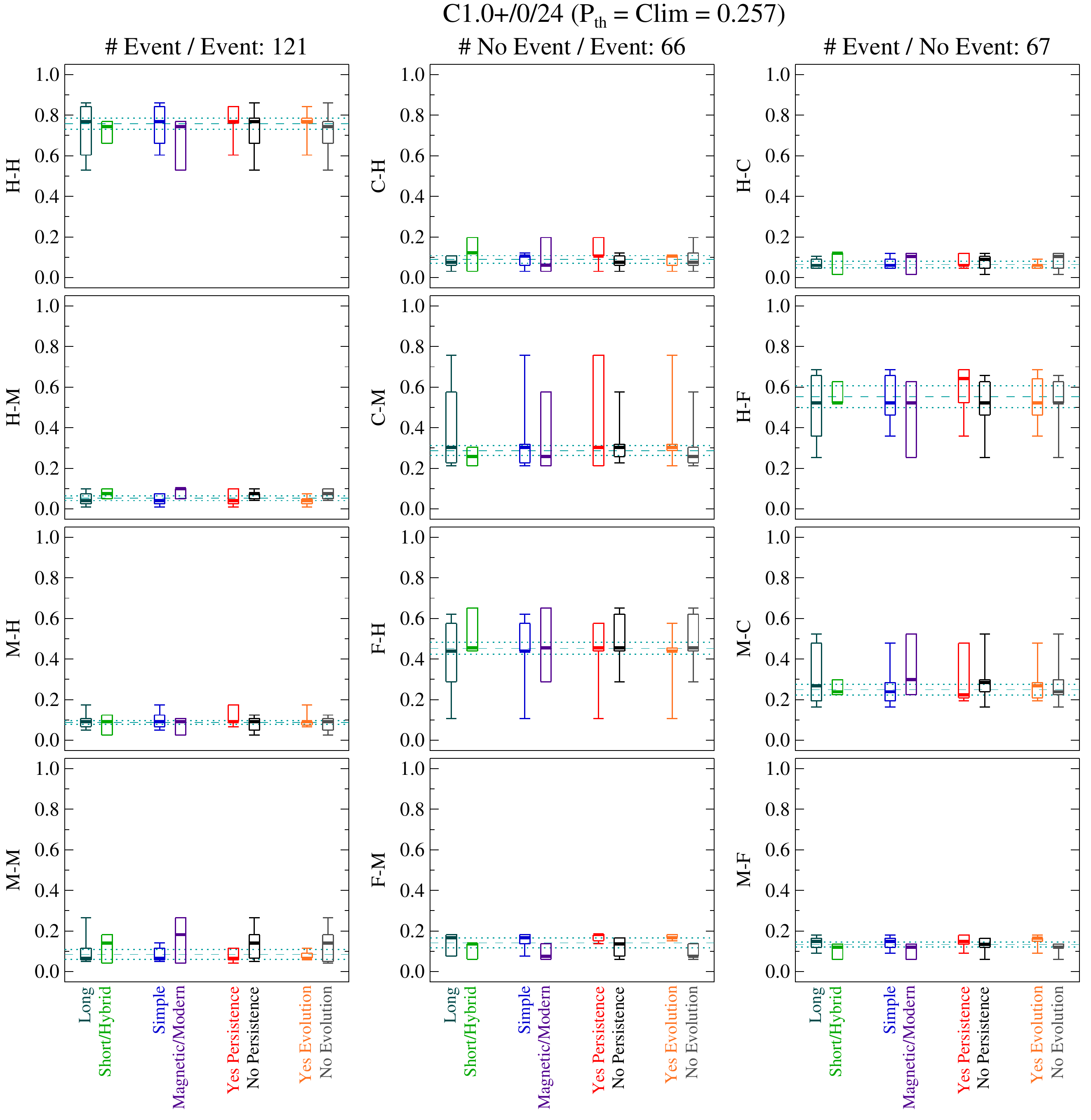}
\caption{Same as in Figure~\ref{fig:box_whisker_c}, but for the \CC\ dichotomous forecasts with $P_\mathrm{th}$\,=\,CLIM.}
\label{fig:box_whisker_c_clim}
\end{figure}

The box and whisker plots in Figures~\ref{fig:box_whisker_m} and~\ref{fig:box_whisker_c} show the relative frequency distribution of the four outcome patterns (rows) for each of the three event histories (columns) for the \MM\ and \CC\ forecasts, respectively, using $P_\mathrm{th}$\,=\,0.5. We describe the results in order of fully correct (top row), fully incorrect (bottom row), and then mixed errors (one of two days correct, one incorrect; middle two rows). Note that this order is according to outcome pattern, whereas the outcome patterns in Table~\ref{tbl:2day} are listed in order according to the forecast made. The evaluations are based here on visual inspection of the medians of the box and whisker plots initially, and secondarily based on the inter-quartile ranges as well. Throughout, we consider the sample size context when considering whether a result is strong or weak.

First, regarding the fully-correct outcome patterns (top row of Figures~\ref{fig:box_whisker_m} and~\ref{fig:box_whisker_c}, where a higher frequency is better), we find differences in the median values between BIOs but at more remarkable magnitudes for the event/event history than for the mixed-event histories. In the \MM\ event/event history, the BIOs of \textit{Yes-Persistence} and \textit{Yes-Evolution} show better performance according to the median, but the performance is similarly improved with the BIOs of \textit{Long} and \textit{Simple}. For the \MM\ mixed-event histories, in most cases the medians and quartiles are very similar, with only small differences in the training and data characterization BIOs for the event/no-event history. For \CC, which has a significantly larger event-day sample size but a smaller number of methods, the BIOs of \textit{Simple} and \textit{Yes-Evolution} consistently show a slightly higher frequency (better performance) according to the medians, but this trend is diluted upon considering the quartile spreads.

Second, we turn to the perfectly-incorrect outcomes (bottom row of Figures~\ref{fig:box_whisker_m} and~\ref{fig:box_whisker_c}, where a lower frequency is better). In the case of the event/event history, there is a mirror effect as compared to the fully-correct outcomes simply due to the lower frequency values of the mixed-error outcome patterns. For the \MM\ event/event history (left column), the \textit{Yes-Persistence} and \textit{Yes-Evolution} BIOs show lower frequency (better performance), however similar differences are found in the median values between the other BIOs, \textit{Long} vs. \textit{Short/Hybrid} and \textit{Simple} vs. \textit{Magnetic/Modern}. For the \MM\ mixed-event histories (middle and right columns), \textit{Yes-Persistence} and \textit{Yes-Evolution} show worse performance according to higher medians of M-F compared to \textit{No-Persistence} and \textit{No-Evolution}, but the same difference is seen between \textit{Long} vs. \textit{Short/Hybrid} as well. For the \CC\ event/event history the medians are effectively the same across all BIOs, but tending toward worsening performance (according to higher 75th percentiles of M-M) for the BIOs that did not include temporal information ({\it e.g.}, \textit{No-Persistence} and \textit{No-Evolution}). For the \CC\ mixed-event histories, we may argue that the \textit{Yes-Persistence} and \textit{Yes-Evolution} BIOs show a worse performance compared to other BIOs, meaning that including temporal information leads to a false alarm after a miss or a miss after a correct null.

Finally, we summarize the mixed-error outcomes for all three event histories (middle two rows of Figures~\ref{fig:box_whisker_m} and~\ref{fig:box_whisker_c}, where again a lower frequency is better). Starting with \MM\ and across the three different event histories, ``first-day-correct/second-day-incorrect'' outcomes (2nd row of Figure~\ref{fig:box_whisker_m}) show many significant differences between BIOs. \textit{Long}, \textit{Simple} and \textit{Yes-Evolution} (but not \textit{Yes-Persistence}) are advantageous for the event/event history, but lead to more second-day errors for the event/no-event history. In the no-event/event history, we see a higher frequency (poorer performance) for \textit{Yes-Persistence} and \textit{Yes-Evolution} compared to the other BIOs. In the case of ``first-day-incorrect/second-day-correct'' patterns (3rd row of Figure~\ref{fig:box_whisker_m}), opposite trends are found in the three event histories. For \CC, there is a tendency that when a significant difference does occur between a pair of BIOs for one of the mixed-event histories, the difference tends to be in the same direction between that pair for the other of the mixed-event histories. An improvement in one ({\it e.g.}, C-M) is reflected in an improvement in the other ({\it e.g.}, H-F). Another finding is that both \textit{Short/Hybrid} and \textit{Magnetic/Modern} consistently show a lower frequency (better performance) with respect to all mixed-error patterns for the mixed-event histories.

In Figures~\ref{fig:box_whisker_m_clim} and~\ref{fig:box_whisker_c_clim}, the same analysis of the BIOs is applied using $P_\mathrm{th}$\,=\,CLIM. In the case of \MM, improvements in performance are found for the event/event history by \textit{Long} and for the event/no-event history by \textit{Short/Hybrid} as well as \textit{Magnetic/Modern} options. \textit{Yes-Persistence} shows a statistically significant increase in the median (better performance) for the fully-correct outcomes C-H and H-C in the mixed-event histories. On the other hand, for \CC\ the first-day-correct/second-day-incorrect outcomes (second row) indicate that \textit{Yes-Persistence} leads to higher errors -- {\it e.g.}, for H-F, acting on the first day's activity leads to over-predict when activity declines on the second day.

\subsection{Targeted Questions Answered}
\label{subsec:results_questions}
We apply nonparametric statistical tests and decision trees (Appendix~\ref{sec:appendix_questions}) to answer the four questions presented in Section~\ref{subsec:intro_questions}. This approach is taken to allow more specific and quantitative analysis to the visual inspection of the box and whisker plots presented above.
      
\paragraph{What is the impact of the BIOs on the independence of the two-day forecasts (meaning, does the forecast outcome for the first day significantly influence the forecast outcome of the second day)?}
For each forecasting method, we test the null hypothesis stated as, ``the forecast outcome on the second day is independent of the forecast outcome of the first day.'' Employing a special contingency table (see Table~\ref{tbl:contingency}) that relates the first- {\it vs.} second-day forecast outcomes, we calculate the significance level, specifically the two-sided $p$-value, from Fisher's exact test \citep{1970smrw.book.....F} of the null hypothesis: lower $p$-values indicate a lower probability of accepting the null hypothesis, {\it i.e.}, a higher likelihood that the day-2 forecast outcome is in fact being influenced by the outcome of the day-1 forecast. The mean of the $p$-values across all forecasting methods in a given BIO is shown in each cell of Table~\ref{tbl:questions_1} as per the event definition, the $P_\mathrm{th}$ value used, and the two-day event history. 

The sample sizes for the \MM\ and \CC\ event definitions are significantly different, leading to the very different magnitudes of $p$-values between the two. As such we only compare relative $p$-values within each definition separately and highlight relatively significant results. 

With \MM\ and $P_\mathrm{th}$\,=\,0.5, we call out \textit{Yes-Evolution} for the event/event history and \textit{Simple} for the event/no-event history as having the two smallest $p$-values but not yet significant at the p=0.05 level, indicating that their forecast outcomes across the two days are less likely to be independent. In Table~\ref{tbl:contingency_example} we present the contingency table entries across all methods in the BIO for these two cases with the expected populations under the null hypothesis shown in parentheses. It is clear with this demonstration that small $p$-values can arise due to over-population of either the on-diagonal or off-diagonal elements.

\begin{table}
\caption{Two-sided $p$-values from Fisher's Exact Test of Independence for Two-day Forecasts}
\advance\tabcolsep-2.4pt
\footnotesize
\begin{center}
\renewcommand{\arraystretch}{1.2}
\begin{tabular}{c|c|c|c|c|c|c|c|c|c|c}
\toprule
\multirow{3}{*}{Event Definition} & \multirow{3}{*}{$P_\mathrm{th}$} & \multirow{3}{*}{Event History} & \multicolumn{2}{c|}{Training Interval} & \multicolumn{2}{c|}{Input Parameter} & \multicolumn{2}{c|}{Persistence} & \multicolumn{2}{c}{Evolution} \\ \cline{4-11}
 & & & \multirow{2}{*}{Long} & Short/ & \multirow{2}{*}{Simple} & Magnetic/ & \multirow{2}{*}{Yes} & \multirow{2}{*}{No} & \multirow{2}{*}{Yes} & \multirow{2}{*}{No} \\ 
 & & & & Hybrid & & Modern & & & & \\ \hline\hline
\multirow{3}{*}{\MM} & \multirow{3}{*}{0.5} & Event / Event & 0.27 & 0.46 & 0.25 & 0.50 & 0.31 & 0.43 & 0.19 & 0.47 \\ \cline{3-11}
 & & No Event / Event & 0.52 & 0.73 & 0.48 & 0.79 & 0.88 & 0.44 & 0.67 & 0.61 \\ \cline{3-11}
 & & Event / No Event & 0.31 & 0.68 & 0.18 & 0.85 & 0.60 & 0.45 & 0.23 & 0.66 \\ \cline{1-11}
\multirow{3}{*}{\CC} & \multirow{3}{*}{0.5} & Event / Event & 2.2$\cdot$10$^{\text{-}6}$ & 4.9$\cdot$10$^{\text{-}6}$ & 2.2$\cdot$10$^{\text{-}6}$ & 4.9$\cdot$10$^{\text{-}6}$ & 4.6$\cdot$10$^{\text{-}6}$ & 2.0$\cdot$10$^{\text{-}6}$ & 3.1$\cdot$10$^{\text{-}6}$ & 2.9$\cdot$10$^{\text{-}6}$ \\ \cline{3-11}
 & & No Event / Event & 6.8$\cdot$10$^{\text{-}3}$ & 3.7$\cdot$10$^{\text{-}3}$ & 2.6$\cdot$10$^{\text{-}3}$ & 0.01 & 6.1$\cdot$10$^{\text{-}3}$ & 5.7$\cdot$10$^{\text{-}3}$ & 3.4$\cdot$10$^{\text{-}3}$ & 8.3$\cdot$10$^{\text{-}3}$ \\ \cline{3-11}
 & & Event / No Event & 6.4$\cdot$10$^{\text{-}3}$ & 4.1$\cdot$10$^{\text{-}3}$ & 2.6$\cdot$10$^{\text{-}5}$ & 0.02 & 3.1$\cdot$10$^{\text{-}3}$ & 7.5$\cdot$10$^{\text{-}3}$ & 3.6$\cdot$10$^{\text{-}6}$ & 0.01 \\ \cline{1-11}
\multirow{3}{*}{\MM} & \multirow{3}{*}{CLIM} & Event / Event & 0.92 & 0.58 & 0.93 & 0.53 & 0.59 & 0.84 & 0.89 & 0.65 \\ \cline{3-11}
 & & No Event / Event & 0.11 & 0.22 & 0.12 & 0.23 & 0.15 & 0.19 & 0.13 & 0.19 \\ \cline{3-11}
 & & Event / No Event & 0.12 & 0.11 & 0.12 & 0.11 & 0.21 & 0.04 & 0.16 & 0.09 \\ \cline{1-11}
\multirow{3}{*}{\CC} & \multirow{3}{*}{CLIM} & Event / Event & 1.4$\cdot$10$^{\text{-}4}$ & 0.02 & 1.4$\cdot$10$^{\text{-}4}$ & 0.02 & 0.01 & 1.6$\cdot$10$^{\text{-}4}$ & 1.8$\cdot$10$^{\text{-}4}$ & 0.01 \\ \cline{3-11}
 & & No Event / Event & 3.7$\cdot$10$^{\text{-}4}$ & 6.7$\cdot$10$^{\text{-}3}$ & 3.9$\cdot$10$^{\text{-}4}$ & 6.7$\cdot$10$^{\text{-}3}$ & 5.6$\cdot$10$^{\text{-}3}$ & 9.1$\cdot$10$^{\text{-}5}$ & 5.1$\cdot$10$^{\text{-}4}$ & 4.0$\cdot$10$^{\text{-}3}$ \\ \cline{3-11}
 & & Event / No Event & 2.7$\cdot$10$^{\text{-}5}$ & 2.0$\cdot$10$^{\text{-}4}$ & 4.2$\cdot$10$^{\text{-}5}$ & 1.7$\cdot$10$^{\text{-}4}$ & 1.2$\cdot$10$^{\text{-}4}$ & 5.5$\cdot$10$^{\text{-}5}$ & 2.4$\cdot$10$^{\text{-}5}$ & 1.4$\cdot$10$^{\text{-}4}$ \\ 
\bottomrule 
\end{tabular}
\renewcommand{\arraystretch}{1}
\end{center}
\tablecomments{The details of the calculation of the $p$-values are described in Appendix~\ref{subsec:appendix_questions_1}.}
\label{tbl:questions_1}
\end{table}

\begin{table}[h!]
\caption{Examples of Contingency Tables for Two-day Forecasts}
\begin{center}
\begin{tabular}{l|l|l|c|c}
\toprule
 & \multicolumn{2}{l|}{} & \multicolumn{2}{c}{First Day Forecast} \\ \cline{4-5}
\textit{Yes-Evolution} (Event / Event) & \multicolumn{2}{l|}{} &  Correct & Incorrect \\ \cline{2-5} 
\MM\ with $P_\mathrm{th}$\,=\,0.5 & Second Day & Correct   & 29 (18.1) & 13 (23.9) \\ \cline{3-5}
$p$\,=\,0.19 & Forecast   & Incorrect & 2 (12.9) & 28 (17.1) \\
\hline\hline
 & \multicolumn{2}{l|}{} & \multicolumn{2}{c}{First Day Forecast} \\ \cline{4-5}
\textit{Simple} (Event / No Event) & \multicolumn{2}{l|}{} &  Correct & Incorrect \\ \cline{2-5}
\MM\ with $P_\mathrm{th}$\,=\,0.5 & Second Day & Correct   & 4 (17.5)  & 98 (84.5)  \\ \cline{3-5}
$p$\,=\,0.18 & Forecast   & Incorrect & 18 (4.5)  & 8 (21.5) \\
\hline\hline
 & \multicolumn{2}{l|}{} & \multicolumn{2}{c}{First Day Forecast} \\ \cline{4-5}
\textit{Long} (Event / Event) & \multicolumn{2}{l|}{} &  Correct & Incorrect \\ \cline{2-5}            
\MM\ with $P_\mathrm{th}$\,=\,CLIM & Second Day & Correct   & 85 (84.3) & 7 (7.7)  \\ \cline{3-5}
$p$\,=\,0.92 & Forecast   & Incorrect & 3 (3.7)   & 1 (0.3) \\
\hline\hline
 & \multicolumn{2}{l|}{} & \multicolumn{2}{c}{First Day Forecast} \\ \cline{4-5}
\textit{No-Persistence} (Event / No Event) & \multicolumn{2}{l|}{} &  Correct & Incorrect \\ \cline{2-5}            
\MM\ with $P_\mathrm{th}$\,=\,CLIM & Second Day & Correct   & 9 (31.2) & 36 (13.8)  \\ \cline{3-5}
$p$\,=\,0.04 & Forecast   & Incorrect & 88 (65.8)   & 7 (29.2) \\
\bottomrule 
\end{tabular}
\end{center}
\label{tbl:contingency_example}
\end{table}

For \MM\ with $P_\mathrm{th}$\,=\,CLIM, large $p$-values are found in the event/event history across all BIOs, due to the fact that the occurrence frequency of H-H is overwhelmingly higher than those of the other three outcome patterns. This is demonstrated in the third set of entries in Table~\ref{tbl:contingency_example}. On the other hand, \textit{No-Persistence} shows a small $p$-value of 0.04 for the event/no-event history due to over-population of off-diagonal elements as shown in Table~\ref{tbl:contingency_example}.

For \CC, extremely small $p$-values (10$^{-6}$--10$^{-2}$) across the BIOs result from the larger sample size coupled with either the on-diagonal or the off-diagonal totals in the contingency tables always being much larger than any of the marginal totals. A tendency is found: for the event/event history either both the day-1 and day-2 forecasts are correct or neither of them is correct, while for the other mixed-event histories only one of the two-day forecasts is correct. This leads us to further test the hypothesis that the day-1 forecast is more likely to be followed by the same day-2 forecast than if the two forecasts were independent of each other. Examining the difference between the contingency table entries and their expected values under the null hypothesis, we find across all BIOs that: for the event/event history the on-diagonal entries (H-H and M-M) consistently exceed their expected values, while for the mixed event histories the off-diagonal entries (F-H and C-M for the no-event/event history and M-C and H-F for the event/no-event history) exceed their expected values. This indicates that a forecast probability higher/lower than $P_\mathrm{th}$ on day-1 tends to stay higher/lower than $P_\mathrm{th}$ on day-2 across all BIOs, as initially identified in Figure~\ref{fig:prob_diagram_m_1_demo}. It seems that forecasts do not respond fast enough ({\it i.e.}, within 24 hours) to changes in the flaring history -- this is a wide-spread failure in forecasting methods, and a specific target for improvement.

\paragraph{Is there any overall performance difference between the BIOs within each particular categorization?}
This is simply ``who wins?'' across all outcome patterns and the implied forecasting performance, between the BIOs -- looking across event definitions and $P_\mathrm{th}$ used. The decision tree associated with this question (see Appendix~\ref{subsec:appendix_questions_2}) is applied; a higher score indicates better performance.

As shown in Table~\ref{tbl:questions_2}, for \MM\ with $P_\mathrm{th}$\,=\,0.5, \textit{Long}, \textit{Simple} and \textit{Yes-Persistence} (but not \textit{Yes-Evolution}) perform slightly better with the difference in the scores against their counterpart BIOs of 0.2--0.3. For \CC\ with $P_\mathrm{th}$\,=\,0.5, their counterpart groups ({\it i.e.}, \textit{Short/Hybrid}, \textit{Magnetic}, \textit{No-Persistence} and even \textit{No-Evolution}) perform relatively much better with the difference in the rank-based scores of 0.5--0.8.

Comparing the performance results across the event definitions and $P_\mathrm{th}$ values used, \textit{No-Evolution} always performs better than \textit{Yes-Evolution} with score differences in the 0.2--1.1 range; \textit{Yes-Persistence} performs better than \textit{No-Persistence} except the case of \CC\ with $P_\mathrm{th}$\,=\,0.5; \textit{Short/Hybrid} performs better than \textit{Long} except the case of \MM\ with $P_\mathrm{th}$\,=\,0.5. Relative to the maximum possible difference, those discussed here are fairly small.

\begin{table}[t!]
\caption{Performance Comparison between Each Pair of BIOs from Same Broad Categorization}
\begin{center}
\begin{tabular}{c|c|c|c|c|c|c|c|c|c}
\toprule
\multirow{3}{*}{Event Definition} & \multirow{3}{*}{$P_\mathrm{th}$} & \multicolumn{2}{c|}{Training Interval} & \multicolumn{2}{c|}{Input Parameter} & \multicolumn{2}{c|}{Persistence} & \multicolumn{2}{c}{Evolution} \\ \cline{3-10}
 & & \multirow{2}{*}{Long} & Short/ & \multirow{2}{*}{Simple} & Magnetic/ & \multirow{2}{*}{Yes} & \multirow{2}{*}{No} & \multirow{2}{*}{Yes} & \multirow{2}{*}{No} \\ 
 & & & Hybrid & & Modern & & & & \\ \hline\hline 
\MM   & 0.5  & 2.0 & 1.7 & 2.3 & 2.0 & 1.7 & 1.5 & 1.9 & 2.1 \\
\CC   & 0.5  & 0.9 & 1.4 & 1.1 & 1.8 & 1.1 & 1.9 & 1.0 & 1.7 \\
\MM   & CLIM & 2.4 & 2.8 & 1.6 & 1.6 & 1.8 & 1.0 & 1.4 & 2.1 \\
\CC   & CLIM & 1.5 & 2.4 & 2.4 & 2.3 & 2.3 & 1.3 & 1.4 & 2.5 \\ \hline
\multicolumn{2}{c|}{Total} & 6.8 & 8.3 & 7.4 & 7.7 & 6.9 & 5.7 & 5.7 & 8.4 \\ 
\bottomrule
\end{tabular}
\end{center}
\tablecomments{The score of the performance comparison in each cell ranges from 0 to 12. The details of the scoring procedure are described in Appendix~\ref{subsec:appendix_questions_2}.}
\label{tbl:questions_2}
\end{table}

\begin{table}[b!]
\caption{Performance Evaluation of BIOs for Mixed-event Histories}
\begin{center}
\begin{tabular}{c|c|c|c|c|c|c|c|c|c}
\toprule
\multirow{3}{*}{Event Definition} & \multirow{3}{*}{$P_\mathrm{th}$} & \multicolumn{2}{c|}{Training Interval} & \multicolumn{2}{c|}{Input Parameter} & \multicolumn{2}{c|}{Persistence} & \multicolumn{2}{c}{Evolution} \\ \cline{3-10}
 & & \multirow{2}{*}{Long} & Short/ & \multirow{2}{*}{Simple} & Magnetic/ & \multirow{2}{*}{Yes} & \multirow{2}{*}{No} & \multirow{2}{*}{Yes} & \multirow{2}{*}{No} \\ 
 & & & Hybrid & & Modern & & & & \\ \hline\hline 
\MM   & 0.5  & 0.5 & 0.9 & 0.5 & 1.2 & 0.9 & 0.7 & 0.5 & 1.0 \\
\CC   & 0.5  & 0.5 & 0.3 & 0.5 & 0.3 & 0.5 & 0.4 & 0.5 & 0.3 \\
\MM   & CLIM & 0.4 & 1.3 & 0.6 & 1.0 & 0.6 & 0.6 & 0.5 & 1.1 \\
\CC   & CLIM & 0.1 & 0.5 & 0.2 & 0.3 & 0.1 & 0.3 & 0.1 & 0.3 \\ \hline
\multicolumn{2}{c|}{Total} & 1.5 & 3.0 & 1.8 & 2.8 & 2.1 & 2.0 & 1.6 & 2.7 \\ 
\bottomrule
\end{tabular}
\end{center}
\tablecomments{The performance score in each cell ranges from 0 to 3; the highest achievable total score is 12. The details of the scoring procedure are described in Appendix~\ref{subsec:appendix_questions_3}.}
\label{tbl:questions_3}
\end{table}

\paragraph{Do any of the BIOs better predict both the first flare and the first quiet?}
This question directly addresses one of the motivations of this study, and we apply the decision tree (Appendix~\ref{subsec:appendix_questions_3}) to achieve the results shown in Table~\ref{tbl:questions_3} where higher totals indicate better performance. We find that \textit{Short/Hybrid}, \textit{Magnetic/Modern} and \textit{No-Evolution} attain higher total scores (2.7--3.0 out of 12), but these scores are equal to or below 25\% of the maximum score possible, hence are not strong results. It is also found that most BIOs show better performance for \MM\ than \CC\ in the context of the first flare/first quiet predictions.

\paragraph{Do those BIOs that explicitly incorporate temporal information ({\it i.e.}, \textit{Yes-Persistence} and \textit{Yes-Evolution}) display performance differences as compared to those BIOs that do not include explicit temporal information?}
To answer this final question we re-examine Tables~\ref{tbl:questions_1}--\ref{tbl:questions_3}. We find that \textit{Yes-Persistence} and \textit{Yes-Evolution} show some overall performance differences compared to the other BIOs but the differences are not large. With respect to having a higher frequency of the two-day-correct patterns as well as a lower frequency of the error patterns, the performance comparison between a pair of BIOs in Table~\ref{tbl:questions_2} shows that \textit{Yes-Persistence} performs better for \MM\ as well as \CC\ with $P_\mathrm{th}$\,=\,CLIM, but \textit{Yes-Evolution} performs worse across all event definitions. In addition, the performance evaluation for the mixed-event histories in Table~\ref{tbl:questions_3} shows similar results as in Table~\ref{tbl:questions_2}. In summary, we find weak support for improvements in performance particularly in the case of \MM\ by explicitly including persistence or prior flare history but excluding active region evolution.

\subsection{Limb Events}
\label{subsec:limb_events}
\begin{figure}[t!]
\centering
\includegraphics[width=0.78\textwidth]{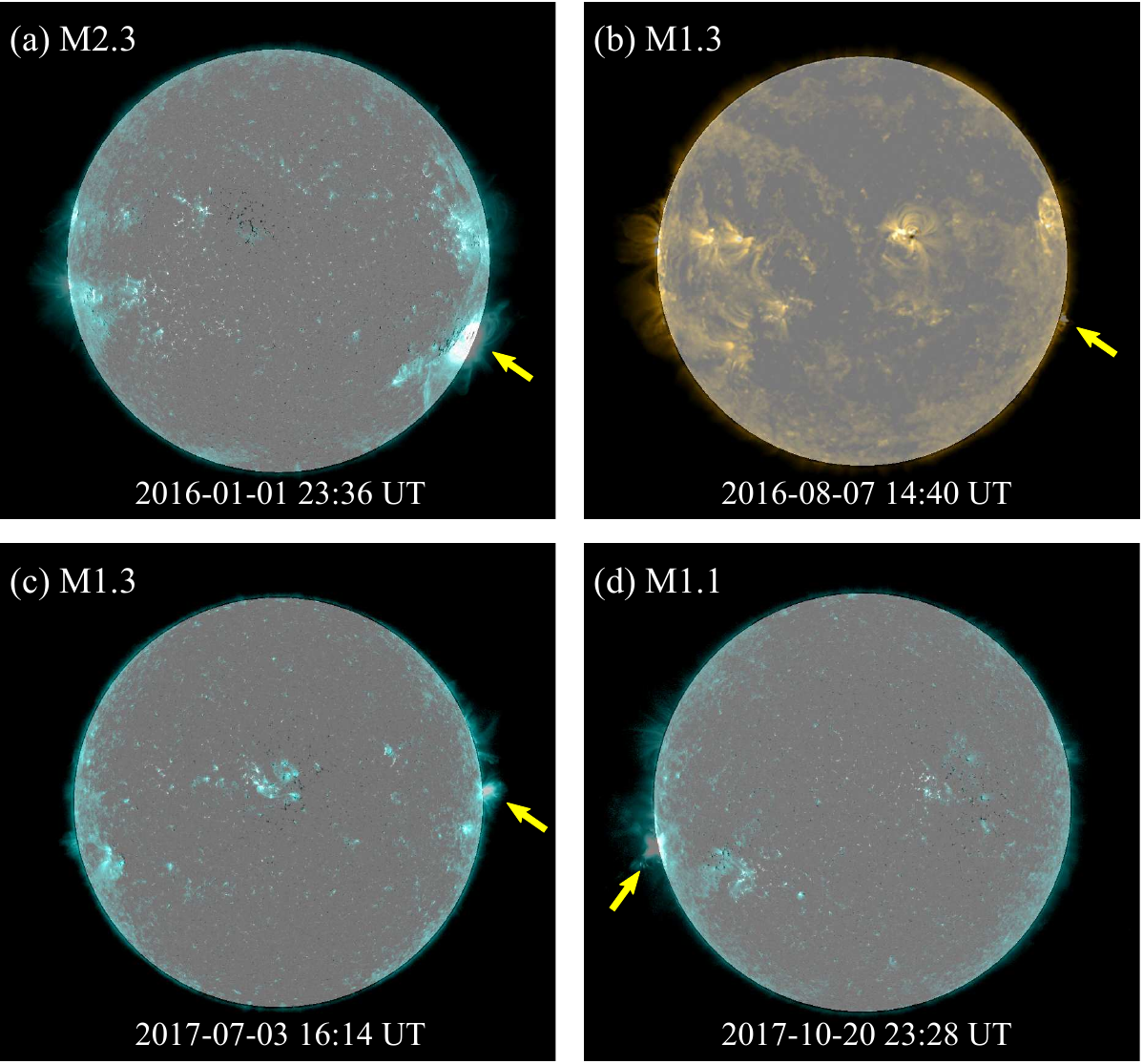}
\caption{Summary images for the four at- or behind-limb large flares (see text). The location of each event is marked by the arrow on a full-disk composite image of the Sun obtained from SDO/HMI line-of-sight magnetic field and SDO/AIA 131\AA\ (or PROBA2/SWAP 174\AA\ in panel (b)). The GOES start times and peak 1--8\AA\ soft X-ray fluxes are also indicated.}
\label{fig:limb_events}
\end{figure}

Finally, breaking of multi-day forecast outcomes, we note that the first flare/first quiet challenges are even more stringent when faced with very isolated flare events, {\it i.e.} when very low activity is interrupted by a single event-day (see Figures~\ref{fig:overview_m} and~\ref{fig:overview_c}). At the $P_\mathrm{th}$\,=\,0.5 level, there are four \MM\ event-days (of 26 event-days, or 15\%) for which all methods failed to provide a ``yes'' forecast (all methods registered a miss). These four event-days share common traits: (1) only one flare event occurred on those days; (2) the source active region was located close to or behind the solar limb; (3) solar activity was very low with few or even no sunspots on the solar disk (see Figure~\ref{fig:limb_events} and Table~\ref{tbl:limb_events}). Examining these four events in some detail provides insight into the possibility of improving the forecasting in these situations.

\begin{table}[b!]
\caption{Summary for Limb Flares on Four Event-days}
\begin{center}
\begin{tabular}{l|c|c|c|c}
\toprule
\multicolumn{3}{c|}{Flare} & \multicolumn{2}{c}{Source Region} \\ \hline
\# &  Start Time &  Peak Flux & NOAA Number & Location \\ \hline 
1 &  2016-01-01 23:10 UT & M2.3 & 12473 & S25 W82 \\
2 &  2016-08-07 14:37 UT & M1.3 & None\tablenotemark{$^\dagger$} & S12 W70\tablenotemark{$^\S$} \\
3 &  2017-07-03 15:37 UT & M1.3 & None\tablenotemark{$^\dagger$} & N02 W85 \\
4 &  2017-10-20 23:10 UT & M1.1 & 12685 & S12 E88\tablenotemark{$^\S$} \\ 
\bottomrule            
\end{tabular}
\end{center}
\tablecomments{Locations and active region assignments from the NOAA Edited Solar Events and Solar Region Summary archive (\url{ftp://ftp.swpc.noaa.gov/pub/warehouse}) and the {\it SolarSoft} Latest Events catalog (\url{https://www.lmsal.com/solarsoft/latest_events}).} \tablenotetext{^\dagger}{No region number assigned before or after with which to associate this flare. See the details in Section~\ref{subsec:limb_events}.} \tablenotetext{^\S}{Longitude uncertain in relevant imaging; likely behind the limb.}
\label{tbl:limb_events}
\end{table}

For Limb Event \#1, AR 12473 had produced {\tt M1.0+} flares a few days prior, but then became quiet. At the $P_\mathrm{th}$\,=\,CLIM the majority of methods forecast \MM\ events to occur, even though for the prior few days it had only produced low-{\tt C}-class events. NJIT solely predicted a significantly higher probability ($>$ 20\%) for \MM\ when this region was at the limb (see Figure~\ref{fig:limb_events}{a}).

In contrast, Limb Event \#2 was an {\tt M1.3} event produced from a fast-growing active region that first appeared close to the western limb. The {\it NOAA Edited Solar Events} archive provides no information regarding location or source region for this flare. However, it was observed by both SDO/AIA and PROBA2/SWAP 174\AA\ (see Figure~\ref{fig:limb_events}{b}), and was not associated with the more flare-productive ARs 12570/12572. In far-side helioseismic maps a region can be detected at the appropriate position a few days later\footnote{See \url{http://jsoc.stanford.edu/data/farside/Composite\_Maps\_JPEG/}}. The majority of methods predicted an event to occur at the \MM\ level for $P_\mathrm{th}$\,=\,CLIM threshold, but not at $P_\mathrm{th}$\,=\,0.5. Any significant full-disk probability was likely due to other visible regions, and {\it not} from the (unassigned) source region itself.

Limb Event \#3 is very similar to Event \#2, in that the source region was a fast-growing emerging flux region that first appeared as it approached the western limb (see Figure~\ref{fig:limb_events}{c}). It also appears a few days later in far-side helioseismic maps. In contrast with Limb Event \#2, in this case no forecast method produced a probability of an \MM\ event even at $P_\mathrm{th}$\,=\,CLIM.
 
The last flare, Limb Event \#4, occurred at the eastern limb before the source region was directly observable. The active region was not large enough to detect in the days prior using presently-available far-side helioseismic maps. The {\it NOAA Solar Region Summary Report} indicated the expected return of AR 12682 which had been a fairly quiet active region on its prior disk appearance. The source region AR 12685 of Limb Event \#4 produced no further flares, and rapidly decayed with no other active region in the vicinity. Only one method, the no-skill CLIM120 forecast, produced a full-disk forecast probability above the test-interval climatology.

In summary, given the small number of \MM\ event-days, a significant fraction were missed by all methods at the $P_\mathrm{th}$\,=\,0.5 level. The majority of methods produced forecasts for an event at the $P_\mathrm{th}$\,=\,CLIM level for two of the limb flares (recognizing that CLIM=0.036 is an extremely low threshold). However, all methods missed forecasting a flare day for the other two limb events even in these almost ideal forecasting conditions ({\it i.e.}, otherwise quite low activity). It is sobering to acknowledge that 15\% of the event-days for larger flares during this 2-year period were effectively beyond any forecast capability we presently have.

\section{Summary and Discussion}
\label{sec:discussion}
Solar flare forecasts from a number of operational facilities worldwide have now been subjected to a set of novel evaluation methods designed to address specific behavior in the face of varying flare activity levels. The questions asked arise from the kind of information targeted in case studies: do the forecasts correctly identify a period of rising or declining activity? And if not, are there particular implementation options being used which exacerbate forecast errors of either kind (misses or false alarms) when a forecasting method is faced with temporally-varying levels of flare activity? The two-year testing interval targeted here ({\it i.e.}, 2016 to 2017) includes a number of distinct periods of flare activity and quiescence, providing a good laboratory for this analysis. The performance characteristics of the forecasting methods under evaluation can be summarized as follows:
\begin{enumerate}
	\item All methods show a trend that a high/low forecast probability on day-1 remains high/low on day-2, regardless of any observed transition between ``flare-quiet'' and ``flare-active'';	
	\item Overall forecast performance is improved for \MM\ when persistence or prior flare history are explicitly included in computing forecasts;
	\item Using magnetic/modern data leads to improvement in catching the first event-day as well as the first no-event-day (for \MM);
	\item These are four \MM\ event-days (of 26 event-days, or 15\%) during the two-year testing interval for which all methods failed to provide a ``yes'' forecast at $P_\mathrm{th}$\,=\,0.5.
\end{enumerate}

In more detail, the forecast outcomes are constructed as dichotomous forecasts by applying a threshold above which the (mostly) probabilistic output of the forecasting methods is taken to be a ``yes'' forecast at the given event definition ({\it e.g.}, \CC\ or \MM\ in this study) or a ``no'' forecast when the probability values are below. We default to $P_\mathrm{th}$\,=\,0.5 as used in the earlier papers in this series, which reflects an intuitive ``50/50'' threshold such that a forecast probability must be above 0.5 ({\it i.e.}, 50\%) to be considered a forecast of an event. However, as discussed in Paper II, for larger magnitude event definitions most probabilistic forecast output is concentrated at low probabilities and some methods never forecast any high probabilities. Hence, for most of the analysis methods developed herein we also present results for $P_\mathrm{th}$\,=\,CLIM ({\it i.e.}, the climatological event rate for the testing interval itself). 

Because a fairly short period of two years is examined, we begin with a simple graphical depiction of the forecast outcomes in light of the full-disk soft X-ray daily maximum output (Sections~\ref{subsec:intro_ataglance} and~\ref{subsec:results_at_a_glance}). It is here that patterns of forecast outcomes emerge when methods address distinct, discrete periods when solar flare activity rises, persists, and declines again over a short time period. Case studies can be useful, but any single case study may not be reflective of a method's performance when confronted with additional such cases.

As such, we refine the analysis to focus on the critical first flare/last flare ({\it i.e.}, first quiet) challenge in the reality of varying flare activity. Considering two-day intervals with varying event history cases we specifically and statistically evaluate forecast performance in the context of increasing activity (no-event followed by event), continuing activity (event followed by event) or declining activity (event followed by no-event) histories. The radar plots presented in Section~\ref{subsec:results_twoday} demonstrate a quick graphical interpretation tool. From these plots, the general results of under-forecasting, significant frequency of misses, and failure to correctly predict the first flare/first quiet are exquisitely clear when a probability threshold of $P_\mathrm{th}$\,=\,0.5 is applied. With $P_\mathrm{th}$\,=\,CLIM the dominant results include general over-forecasting, high rates of false alarms, but almost no failures on continuing-activity periods. Some methods do show an ability to correctly forecast either the first day only or the second day only, but very few methods show any ability to correctly forecast both days in the mixed-event histories ({\it i.e.}, when flare activity is changing). Many methods show similar patterns to each other due to their similar approaches (previously discussed in Papers II and III).

We apply the broad categorization analysis developed in Paper III to the frequency analysis of the two-day event histories and outcome patterns in order to investigate the reasons behind certain patterns of success or failure. Box and whisker plots (Section~\ref{subsec:results_twodayplus}) identify two-day forecasting patterns that can be interpreted in the context of the different implementations. The higher frequency of M-H when methods include persistence or prior flare history is consistent with a forecast ``adjustment'' for the second of the two days according to the event history of the first day. This can also explain some of the two-day-incorrect patterns ({\it i.e.}, F-M and M-F) as the influence of persistence or prior flare history brings the forecasts out of step with shorter time-scale changes in flare activity.

We additionally ask targeted questions regarding binary implementation options (BIOs) and their effects on the independence of two-day forecasts and their performance (Section~\ref{subsec:results_questions}) with emphasis on successfully predicting both the first flare and first quiet (last flare). Nonparametric statistical tests and decision-tree games were formulated to ingest the input first presented in the box and whisker analysis and provide quantitative answers thus. We first identify that for \CC\ the day-2 forecast outcome is significantly affected by the day-1 forecast outcome across all BIOs in the way that the two-day forecast probabilities tend to remain either higher or lower than than $P_\mathrm{th}$ for the two-day period regardless of changes in the flaring history. We confirm weak support that including persistence or prior flare activity, as well as excluding active region evolution, improves the \MM\ forecasts across all three two-day event histories (Table~\ref{tbl:questions_2}). On the other hand, there is evidence for improved performance of the \MM\ forecasts when flaring activity is transitioning with the use of \textit{magnetic/modern} data, even if it requires a shorter training interval (Table~\ref{tbl:questions_3}).

Except in a few cases, the results of the BIO-based analyses are not definitive. While small sample size is of course one culprit, another reason (as discussed in Paper III) is that the BIOs are not completely independent. As an example, the methods from regional warning centers (NOAA, SIDC, NICT, MOSWOC) all employ human forecasters ({\it i.e.}, FITL) and by extension all use long training series, simple data input, but also all include persistence and active region evolution in their forecasts, even if in a qualitative manner. As a result, differentiation between overlapping methods in different BIOs is diluted by the lack of true control groups where only one BIO is modified at a time.

Forecasting for flare events that occur at or behind the solar limb is known to be problematic. During the two-year testing period here, four {\tt M1.0+} limb events were completely missed by all methods (at the 50\% probability level); two of these events were correctly predicted by the majority of methods, but only at the $P_\mathrm{th}$\,=\,CLIM=0.037 level. For the other two events, all methods completely failed to produce an ``event'' forecast except one instance of a correct $P_\mathrm{th}$\,=\,CLIM=0.037 event forecast from the full-disk ``no-skill'' 120-day climatology method. To summarize, four of 26 {\tt M1.0+} event-days in our two-year sample were missed essentially due to lack of operationally-available observations away from the Earth-Sun line.

We present here new analysis methods by which to evaluate both existing operational forecasting systems as well as research and development phases of systems yet to be deployed. Specific challenges have now been presented for the flare-forecasting research community beyond simply improving metrics such as those presented in Papers II and III. All operational forecasting methods evaluated here fail to respond adequately to changes in flaring activity. As has been acknowledged in the research community \citep[{\it c.f.},][]{2016SoPh..291..411B}, targeted efforts are needed to specifically improve forecast performance over short-term variations in solar flare activity.

\acknowledgments
We wish to acknowledge funding from the Institute for Space-Earth Environmental Research, Nagoya University for supporting the workshop and its participants. We would also like to acknowledge the ``big picture'' perspective brought by Dr. M. Leila Mays during her participation in the workshop. S.-H.P. gratefully acknowledges Dr. Ju Jing for maintaining the NJIT flare forecasting system and providing the archive forecasts. KDL and GB acknowledge that the DAFFS and DAFFS-G tools were developed under NOAA SBIR contracts WC-133R-13-CN-0079 (Phase-I) and WC-133R-14-CN-0103 (Phase-II) with additional support from Lockheed-Martin Space Systems contract \#4103056734 for Solar-B FPP Phase E support. A.E.McC. was supported by an Irish Research Council Government of Ireland Postgraduate Scholarship. M.K.G acknowledges research performed under the A-EFFort project and subsequent service implementation, supported under ESA Contract number 4000111994/14/D/MPR. D.S.B. and M.K.G. were supported by the European Union Horizon 2020 research and innovation programme under grant agreement No. 640216 (FLARECAST project; {\tt http://flarecast.eu}). S.A.M. is supported by the Irish Research Council Postdoctoral Fellowship Programme and the US Air Force Office of Scientific Research award FA9550-17-1-039. The operational Space Weather services of ROB/SIDC are partially funded through the STCE, a collaborative framework funded by the Belgian Science Policy Office.

\vspace{5mm}
\facilities{GONG, GOES (XRS), PROBA2 (SWAP), SDO (AIA, HMI)}

\bibliographystyle{aasjournal}
\bibliography{ref.bib}

\appendix
\section{Participating Methods and Facilities}
\label{sec:appendix_method_table}
In Table~\ref{tbl:methods} we reproduce an abbreviated version of Table 1 from Paper II, listing the methods and facilities involved with this work and the monikers used to refer to them.

\begin{table}[h!]
\caption{Participating Operational Forecasting Methods (Alphabetical by Label Used)}
\advance\tabcolsep-0.8pt
\footnotesize
\begin{center}
\begin{tabular}{p{5.9cm}p{6.1cm}p{1.8cm}p{3.8cm}}
 Institution & Method/Code Name\tablenotemark{$^\dagger$} & Label & Reference(s) \\ \hline \hline
 \multirow{2}{*}{ESA/SSA A-EFFORT Service} & Athens Effective Solar Flare & \multirow{2}{*}{A-EFFORT}  & \multirow{2}{*}{\citet{2007ApJ...661L.109G}} \\ 
 & Forcasting & & \\ \hline
 Korean Meteorological Administration & Automatic McIntosh-based Occurrence & \multirow{2}{*}{AMOS} & \multirow{2}{*}{\citet{2012SoPh..281..639L}} \\
 \& Kyung Hee University (Korea) & probability of Solar activity & & \\ \hline
 University of Bradford (UK) & Automated Solar Activity Prediction & ASAP & \citet{ColakQahwaji2008,ColakQahwaji2009} \\ \hline
 \multirow{2}{*}{Korean Space Weather Center} & \multirow{2}{*}{Automatic Solar Synoptic Analyzer} & \multirow{2}{*}{ASSA} & \citet{ASSA}, \\
 & & & \citet{ASSA_DOC} \\ \hline
 Bureau of Meteorology (Australia) & FlarecastII & BOM  &\citet{Steward_etal_2011,Steward_etal_2017}\\ \hline
 120-day No-Skill Forecast & Constructed from NOAA event lists & CLIM120  & \citet{SharpeMurray2017} \\ \hline
 \multirow{3}{*}{NorthWest Research Associates (US)} & Discriminant Analysis Flare & \multirow{2}{*}{DAFFS} & \multirow{3}{*}{\citet{2018JSWSC...8A..25L}} \\
 & Forecasting System & & \\ \cline{2-3}
 & GONG+GOES only & DAFFS-G & \\ \hline
 & MAG4 (+according to & MAG4W & \\ \cline{3-3}
 NASA/Marshall Space Flight Center & magnetogram source  & MAG4WF & \citet{2011SpWea...9.4003F}, \\ \cline{3-3}
 (US) & and flare-history  & MAG4VW & Appendix A in Paper II \\ \cline{3-3}
 & inclusion) & MAG4VWF & \\ \hline
 \multirow{3}{*}{Trinity College Dublin (Ireland)} & SolarMonitor.org Flare Prediction & \multirow{2}{*}{MCSTAT} & \citet{2002SoPh..209..171G} \\
 & System (FPS) & & \citet{2012ApJ...747L..41B} \\ \cline{2-4}
 & FPS with evolutionary history & MCEVOL  & \citet{2018JSWSC...8A..34M} \\ \hline
 \multirow{2}{*}{Met Office (UK)} & Met Office Space Weather Operations & \multirow{2}{*}{MOSWOC} & \multirow{2}{*}{\citet{2017SpWea..15..577M}} \\ 
 & Center human-edited forecasts & & \\ \hline
 National Institute of Information and & \multirow{2}{*}{NICT-human} & \multirow{2}{*}{NICT} & \multirow{2}{*}{\citet{2017JSWSC...7A..20K}} \\ 
 Communications Technology (Japan) & & & \\ \hline
 New Jersey Institute of Technology (US) & NJIT-helicity & NJIT  & \citet{Park_Chae_Wang_2010} \\ \hline
 NOAA/Space Weather Prediction & & \multirow{2}{*}{NOAA} & \multirow{2}{*}{\citet{Crown2012}} \\
 Center (US) & & & \\ \hline
 \multirow{2}{*}{Royal Observatory of Belgium} & Solar Influences Data Analysis Center & \multirow{2}{*}{SIDC} & \citet{Berghmans_etal_2005}, \\ 
 & human-generated & & \citet{Devos_etal_2014} \\ \hline
\end{tabular}
\end{center}
\tablenotetext{\dagger}{If applicable}
\label{tbl:methods}
\end{table}

\section{Targeted Analysis of the Binary Implementation Options (BIOs)}
\label{sec:appendix_questions}
In this section we delineate the targeted questions posed in Section~\ref{subsec:intro_questions} and describe the analysis method applied. For the targeted questions 2 and 3 specifically, the analysis takes the form of a decision-tree ``game'' by which credit is applied according to a binary ``win/loss'' outcome, as described below.

\subsection{Targeted Question 1}
\label{subsec:appendix_questions_1}
\paragraph{What is the impact of the BIOs on the independence of the two-day forecasts (meaning, does the forecast outcome for the first day significantly influence the forecast made for the second day)?}  
To answer this, we formulate a test of the null hypothesis: the forecast outcome on day-2 is independent of the forecast outcome on day-1. In other words, given the overall frequencies of success on day-1 and on day-2, does the frequency of success on day-2 depend on what was forecast and what occurred on day-1? To test this null hypothesis, we use Fisher's exact test on a 2$\times$2 contingency table constructed as shown in Table~\ref{tbl:contingency}. $a$, $b$, $c$, and $d$ correspond to the occurrence frequencies of the four outcome patterns in each event history as follows: H-H, M-H, H-M, and M-M for the event/event history, C-H, F-H, C-M, and F-M for the no-event/event history, H-C, M-C, H-F, and M-F for the event/no-event history. This test assumes that the marginal totals ({\it i.e.}, $a$+$b$, $c$+$d$, $a$+$c$, $b$+$d$) are held constant. The Fisher's exact test gives the probability of getting the observed contingency table (or a more extreme case) under the null hypothesis. A two-sided $p$-value is derived from this test for each forecast method as per the event definition and the two-day event history. The mean of the $p$-values across all methods in a given BIO is shown in each cell of Table~\ref{tbl:questions_1}.

\begin{table}[h!]
\caption{Contingency Table}
\begin{center}
\begin{tabular}{l|l|c|c}
\toprule
\multicolumn{2}{c|}{} & \multicolumn{2}{c}{First Day Forecast} \\ \hline
\multicolumn{2}{c|}{} &  Correct & Incorrect \\ \hline            
Second Day & Correct   & a & b \\ \cline{2-4}
Forecast   & Incorrect & c & d \\
\bottomrule            
\end{tabular}
\end{center}
\label{tbl:contingency}
\end{table}

\subsection{Targeted Question 2}
\label{subsec:appendix_questions_2}
\paragraph{Is there any overall performance difference between the BIOs within each particular categorization?} 
To answer the question, each pair of BIOs from the same categorization are compared directly. The relative performance in this context means having a higher frequency of the two-day-correct patterns as well as a lower frequency of the three error patterns. For this question, we only compare two BIOs directly, and do not comment on their overall performance, only their relative performance. As such, we adopt a rank-sum approach applied to the frequency of forecast outcomes. The relative performance is then measured using the Mann-Whitney-Wilcoxon (MWW) rank-sum test, a nonparametric statistical test of the difference between two independent samples \citep{Mann1947}. The MWW test involves the calculation of a statistic (called $U$) for the two samples, respectively:
\begin{equation}
U_{x} = R_{x} - \frac{n_{x}(n_{x}+1)}{2}
\end{equation}
where $x$ represents the particular sample, $n_{x}$ and $R_{x}$ indicates the sample size and the sum of ranks respectively. The absolute value of the difference between the $U$ values for the two samples, {\it i.e.}, $\Delta U$, is then used to measure the significance of the difference between the two samples. Normalization of $\Delta U$ to a range $[0,1]$ is achieved by dividing it by its maximum possible value for the given sample sizes ({\it i.e.}, the product of the two sample sizes). Note that the maximum value of $\Delta U$ is derived from the extreme case that the two samples are completely separated ({\it e.g.}, all values from the first sample are less than all values from the second, or {\it vice versa}). The rank-sum analysis is performed method-by-method on the frequency of the forecast outcomes, where the two samples are the two BIOs being compared.

The comparison and scoring are then carried out for each event definition, each $P_\mathrm{th}$ value used, and each BIO pair of the categories, as follows.
\begin{enumerate}
\item For each of the three two-day-correct patterns ({\it i.e.}, H-H, C-H, H-C), we calculate the $U$ values for the BIO pair ({\it e.g.}, A and B), and compare them ({\it i.e.}, $U_\mathrm{A}$ and $U_\mathrm{B}$):
  \begin{enumerate}
  \item If $U_\mathrm{A}$\,$>$\,$U_\mathrm{B}$, then A will get a score of ($|U_\mathrm{A} - U_\mathrm{B}|$)/($n_\mathrm{A}n_\mathrm{B}$), while B will get no score, and {\it vice versa}. Next, go to step 2.
  \item If $U_\mathrm{A}$\,$=$\,$U_\mathrm{B}$, then both A and B will get no score. Next, go to step 2.
  \end{enumerate}
\item For the other nine error patterns ({\it i.e.}, only first-day-correct, only second-day-correct, or all two-day-incorrect), the opposite rule is applied:
  \begin{enumerate}
  \item If $U_\mathrm{A}$\,$<$\,$U_\mathrm{B}$, then A will get a score of ($|U_\mathrm{A} - U_\mathrm{B}|$)/($n_\mathrm{A}n_\mathrm{B}$), while B will get no score, and {\it vice versa}. Next, go to step 3.
  \item If $U_\mathrm{A}$\,$=$\,$U_\mathrm{B}$, then both A and B will get no score. Next, go to step 3. 
  \end{enumerate}
\item Add all scores for the performance comparisons of all 12 patterns, which is shown in each cell of Table~\ref{tbl:questions_2} as per the event definition, the $P_\mathrm{th}$ value used, and the BIO.
\end{enumerate}
In Table~\ref{tbl:questions_2}, the score of the performance comparison in each cell ranges from 0 to 12.

\subsection{Targeted Question 3}
\label{subsec:appendix_questions_3}
\paragraph{Do any of the BIOs better predict both the first flare and first quiet} 
This question is answered through comparisons between two different outcome patterns for the mixed-event histories only ({\it i.e.}, C-H vs F-M, F-H vs C-M for the no-event/event history, H-C vs M-F, M-C vs H-F for the event/no-event history, as discussed in Section~\ref{subsec:intro_twoday}). The MWW rank-sum test is used for the comparison as described in~\ref{subsec:appendix_questions_2}. The rules of the comparison and performance evaluation for a given BIO are as follows.
\begin{enumerate}
\item For the no-event/event history, the comparison of C-H vs F-M as well as F-H vs C-M is carried out:
  \begin{enumerate}
  \item The $U$ values ({\it i.e.}, $U_\mathrm{C\mbox{-}H}$ and $U_\mathrm{F\mbox{-}M}$) of C-H and F-M are compared:
    \begin{enumerate}
    \item If a given BIO has $U_\mathrm{C\mbox{-}H}$\,$>$\,$U_\mathrm{F\mbox{-}M}$, then it will get a score of ($|U_\mathrm{C\mbox{-}H} - U_\mathrm{F\mbox{-}M}|$)/($n_\mathrm{C\mbox{-}H}\,n_\mathrm{F\mbox{-}M}$). Next, go to step 1.b.
    \item If $U_\mathrm{C\mbox{-}H}$\,$\leq$\,$U_\mathrm{F\mbox{-}M}$, then it will get no score. Next, go to step 1.b.
    \end{enumerate}
  \item $U_\mathrm{F\mbox{-}H}$ and $U_\mathrm{C\mbox{-}M}$ are compared:  
    \begin{enumerate}
    \item If $U_\mathrm{F\mbox{-}H}$\,$>$\,$U_\mathrm{C\mbox{-}M}$, then it will get a score of 0.5\,$\times$\,($|U_\mathrm{F\mbox{-}H} - U_\mathrm{C\mbox{-}M}|$)/($n_\mathrm{F\mbox{-}H}\,n_\mathrm{C\mbox{-}M}$). Next, go to step 2.
    \item If $U_\mathrm{F\mbox{-}H}$\,$\leq$\,$U_\mathrm{C\mbox{-}M}$, then it will get no score. Next, go to step 2.
    \end{enumerate}
  \end{enumerate}
\item For the event/no-event history, the comparison of H-C vs M-F as well as M-C vs H-F is carried out analogously:
  \begin{enumerate}
  \item $U_\mathrm{H\mbox{-}C}$\,$>$\,$U_\mathrm{M\mbox{-}F}$ are compared:
    \begin{enumerate}
    \item If $U_\mathrm{H\mbox{-}C}$\,$>$\,$U_\mathrm{M\mbox{-}F}$, then it will get a score of ($|U_\mathrm{H\mbox{-}C} - U_\mathrm{M\mbox{-}F}|$)/($n_\mathrm{H\mbox{-}C}\,n_\mathrm{M\mbox{-}F}$). Next, go to step 2.b.
    \item If $U_\mathrm{H\mbox{-}C}$\,$\leq$\,$U_\mathrm{M\mbox{-}F}$, then it will get no score. Next, go to step 2.b.
    \end{enumerate}
  \item $U_\mathrm{M\mbox{-}C}$,$>$\,$U_\mathrm{H\mbox{-}F}$ are compared:  
    \begin{enumerate}
    \item If $U_\mathrm{M\mbox{-}C}$\,$>$\,$U_\mathrm{H\mbox{-}F}$, then it will get a score of 0.5\,$\times$\,($|U_\mathrm{M\mbox{-}C} - U_\mathrm{H\mbox{-}F}|$)/($n_\mathrm{M\mbox{-}C}\,n_\mathrm{H\mbox{-}F}$).
    \item If $U_\mathrm{M\mbox{-}C}$\,$\leq$\,$U_\mathrm{H\mbox{-}F}$, then it will get no score.
    \end{enumerate}
  \end{enumerate}
\end{enumerate}
In Table~\ref{tbl:questions_3}, the performance score in each cell ranges from 0 to 3.

\subsection{Targeted Question 4}
\label{subsec:appendix_questions_4}
\paragraph{Do those BIOs which explicitly incorporate temporal information ({\it i.e.}, \textit{Yes-Persistence} and \textit{Yes-Evolution}) display performance differences as compared to those BIOs which do not include explicit temporal information?}
There is no separate statistical test or decision tree needed to address this question. We answer by examining Tables~\ref{tbl:questions_1}--\ref{tbl:questions_3} overall, and in particular, comparing \textit{Yes-Persistence} and \textit{Yes-Evolution} outcomes versus \textit{No-Persistence} and \textit{No-Evolution}, and all other BIOs across the three prior questions.

\end{document}